\documentclass{aa}  

\usepackage{graphicx}
\usepackage{txfonts}
\usepackage{amsmath}
\usepackage[dvipsnames]{xcolor}

\usepackage[colorlinks=true, allcolors=blue]{hyperref}
\newcommand{\msun}{\mathrm{M_\odot}}
\newcommand{\rfh}{\mathrm{R}_{500}}
\newcommand{\rth}{\mathrm{R}_{200}}
\newcommand{\leff}{l_{\mathrm{eff}}}
\newcommand{\tszx}{\mathrm{T_{SZX}}}
\newcommand{\tmw}{\mathrm{T_{mw}}}
\newcommand{\tsl}{\mathrm{T_{sl}}}
\newcommand{\effP}{\overline{P_e}}
\newcommand{\effn}{\overline{n_e}}
\newcommand{\diffd}{\mathrm{d}}

\defcitealias{2017A&A...606A..64A}{A17}

\begin{document}

   \title{Galaxy cluster temperature maps from joint X-ray and SZ maps with The Three Hundred hydrodynamical simulations}
    

   \author{R. Wicker\inst{1}
          \and
          M. De Petris\inst{1,2}
          \and 
          A. Ferragamo\inst{3}
          \and 
          I. Bartalucci\inst{4}
          \and 
          G.Yepes\inst{5,6}
          \and
          E. Rasia\inst{7,8,9}
          \and
          R. Adam\inst{10}
          \and
          W. Cui\inst{5,6,11}
          \and
          F. Mayet\inst{12}
          \and
          L. Perotto\inst{12}
          \and
          M.~Mu\~noz-Echeverr\'ia\inst{13}
          }

   \institute{Dipartimento di Fisica, Sapienza Università di Roma, Piazzale Aldo Moro 5, I-00185 Roma, Italy\\
              \email{raphael.wicker@uniroma1.it}
        \and
        INAF-Osservatorio Astronomico di Roma, Via Frascati 33, I-00078 Monteporzio Catone, Italy      
        \and
        Physics Department “Ettore Pancini”, Università degli studi di Napoli “Federico II”, Via Cintia 21, I-80126 Napoli, Italy
        \and
        INAF/IASF-Milano, Via A. Corti 12, 20133 Milano, Italy
        \and
        Departamento de Física Teórica, Facultad de Ciencias, Universidad Autónoma de Madrid, 28049 Cantoblanco, Madrid, Spain
        \and
        Centro de Investigación Avanzada en Física Fundamental (CIAFF),
        Facultad de Ciencias, Universidad Autónoma de Madrid, 28049 Madrid, Spain
        \and
        INAF-Osservatorio Astronomico di Trieste, via Tiepolo 11, I-34131, Trieste, Italy
        \and
        IFPU-Institute for Fundamental Physics of the Universe, via Beirut 2, 34151, Trieste, Italy
        \and
        Department of Physics, University of Michigan, 450 Church St, Ann Arbor, MI 48109, USA
        \and
        Université Côte d'Azur, Observatoire de la Côte d’Azur, CNRS, Laboratoire Lagrange, France
        \and
        Institute for Astronomy, University of Edinburgh, Royal Observatory, Blackford Hill, Edinburgh EH9 3HJ, UK
        \and
        Univ. Grenoble Alpes, CNRS, Grenoble INP, LPSC-IN2P3, 53, avenue des Martyrs, 38000 Grenoble, France
        \and
        IRAP, CNRS, Université de Toulouse, CNES, UT3-UPS, (Toulouse), France 
        }

   \date{Received XXX; accepted XXX}

 
  \abstract
   {Galaxy clusters can be used as powerful cosmological probes, provided one can obtain accurate mass estimates, which requires a precise knowledge of the underlying astrophysics of galaxy clusters. For these purposes, spatially resolved measurements of the thermodynamic properties of intra-cluster medium (ICM), such as density and temperature, are necessary. In particular, temperature estimates are traditionally obtained through spatially resolved X-ray spectroscopy. Such measurements suffer from their sensitivity to the chosen energy calibration, may exhibit inherent biases, and are especially hard to perform at high redshift as they require deep observations. In recent years however, millimetre wavelength data with high spatial resolution, comparable to the one of current X-ray telescopes, have begun to be available. This has enabled the implementation of new methods to infer and map the cluster temperature in individual clusters, using the combination of density maps from X-ray data and pressure maps from millimetre data. In this paper, we present the first systematic validation of this approach on a large sample of synthetic clusters generated in The Three Hundred hydrodynamical simulations. We show that we are able to recover theoretical estimates of the temperature, namely the mass-weighted and spectroscopic-like temperatures, within biases of the order of $\lesssim 1\%$ in the best cases, up to $\sim 10\%$ in average, with scatters of the order of $10\%$.
   To prepare the application of this approach to observed data, we discuss the modelling of the effective length $\leff$, a key quantity necessary for the combination of X-ray and SZ projected data. In particular we provide templates calibrated on simulations for this quantity, and investigate their impact in the recovery of the temperature map, compared to other standard models.}

   \keywords{Galaxies: clusters: intracluster medium -- X-rays: galaxies: clusters -- SZ effect: galaxies: clusters -- Methods: numerical -- Methods: statistical
               }
   \maketitle
%

\section{Introduction}

Sitting at the nodes of the cosmic web as massive, virialized structures, galaxy clusters can be used as robust cosmological probes \citep{1993Natur.366..429W,2011ARA&A..49..409A}, notably due to the dependence of the Halo Mass Function (HMF) on the geometry of the universe.
However, the quality of the cluster-based cosmological constraints is extremely dependent on the mass calibration \citep{2019SSRv..215...25P}, and as such accurate measurements of the galaxy cluster masses are crucial.

The most common estimates of the cluster mass using observations of the gas content, i.e. using the X-ray emission of the Intra-Cluster Medium \citep[ICM,][]{1988xrec.book.....S} or relying on the Sunyaev-Zel'dovich effect \citep[SZ, ][]{1972CoASP...4..173S}, are carried out under the assumption of the hydrostatic equilibrium (HE).
However, a variety of astrophysical effects and non thermal processes in the gas, such as turbulence, bulk motions, shocks, magnetic fields or cosmic rays \citep{2009ApJ...705.1129L,2009A&A...504...33V,2012ApJ...758...74B,2014ApJ...792...25N,2015MNRAS.448.1020S,2016ApJ...827..112B,Gianfagna2023}, may be the source of departures from the ideal conditions of HE.
The resulting mass estimates will in consequence be underestimated \citep[see e.g.][]{2006MNRAS.369.2013R,2012NJPh...14e5018R,2021MNRAS.502.5115G,2023A&A...674A..48W,2024A&A...682A.147M}, with a strong impact on the subsequent cosmological constraints \citep[see the reviews by][]{2005RvMP...77..207V, 2019SSRv..215...25P}.
As the estimation of the hydrostatic mass relies on using radial profiles of the gas density, gas pressure, and gas temperature, a deep understanding and characterisation of the ICM and its thermodynamical properties are thus critical.
In that respect, temperature measurements of the ICM are of particular interest and importance. 
Indeed, 1D temperature profiles are necessary when computing the hydrostatic mass of a cluster from X-ray data. 
If in addition one wants to study the astrophysical effects in the gas including turbulence, gas sloshing, or shocks, a 2D mapping of the temperature allows to access the spatial distribution of the gas temperature, providing invaluable insight into the ICM physics \citep{2007PhR...443....1M}. 

The most standard estimate of the cluster gas temperature in observations is obtained using spatially resolved X-ray spectroscopy.
Albeit widely used, these measurements are however plagued by numerous limitations.
First, the X-ray flux is heavily dependent on redshift, decreasing as $(1+z)^{-4}$, making high $z$ observations extremely costly in terms of integration time, with objects appearing extremely faint.
As a result, a 2D mapping of the temperature in high redshift clusters proves to be extremely challenging, especially to carry out in a systematic manner.
In addition, X-ray estimates of the temperature will be affected by systematics of various sources.
On one side, the X-ray signal is sensitive to the squared gas density, which naturally leads X-ray observations to probe preferentially only the denser, colder, regions of the core. 
As a result, the temperature measured through spectroscopy will actually be a weighted mean of the temperature along the line of sight, where the weights are a non linear combination of the gas density and temperature \citep{2004MNRAS.354...10M,2006ApJ...640..691V,2014MNRAS.439..588B}.
Another source of systematics stems from the fact that the X-ray spectroscopy and the sub-sequent temperature estimates are extremely dependent on the energy calibration of the instruments used in the observations. 
This will lead the temperature estimates of different X-ray observatories to show strong discrepancies between one another \citep{2015A&A...575A..30S,2024A&A...688A.107M}.
These disagreements may indeed go up to 15\% above 10keV between {\it Chandra} and XMM-{\it Newton} \citep{2015A&A...575A..30S}, and even above 30\% at $\sim 10$keV between SRG/eROSITA and the two aformentioned observatories \citep{2024A&A...688A.107M}.
Finally, the uncertainties in temperature measurements increase with the temperature. 
As a result, the X-ray estimates of the ICM temperatures, notably with XMM-{\it Newton} and {\it Chandra} are especially hard to obtain at high temperatures.

As a result of these shortcomings, alternative methods to infer galaxy cluster temperatures need to be proposed and investigated. 
In order to infer 1D temperature profiles, or even directly mass profiles, works like those of \cite{2007MNRAS.382..397A, 2009MNRAS.394..479A,Mroczkowski_2009,2015A&A...576A..12A,2016A&A...586A.122A,2017A&A...597A.110R,2018A&A...614A...7G} have started using a combination of X-ray and SZ observations, using the density information from the former and the pressure information from the latter.

\citet{2017A&A...606A..64A} (which we will note \citetalias{2017A&A...606A..64A} in the rest of the paper) went a step further and used a combination of XMM-{\it Newton} and high angular resolution NIKA SZ data to infer a map of the temperature in the cluster MACS J0717.5+3745.
Similarly, Artis et al. (in prep) combine XMM-{\it Newton} and NIKA2 observations and manage to highlight an overheated region in the cluster PSZ2G091.83+26.11.

The purpose of this work, built as a series of two papers, is to validate this approach on a large simulated sample of clusters, in order to then apply it in a systematic manner to an observed sample of galaxy clusters, for which both X-ray imagery and high resolution SZ imagery are available, namely the NIKA2 SZ Large Program \citep{2020EPJWC.22800017M,2022EPJWC.25700038P}.
The current paper, first of the series, is focused on the theoretical work in a simulated setup, while the second paper will focus on the application of the method to realistic mock maps and to observed data.
In Section \ref{sect:Data} we detail our simulated dataset and the maps we use for this work. 
The full theoretical modelling necessary to the approach is detailed in Section \ref{sect:theo_model}, while we detail our results in Section \ref{sect:results}, with a focus set on the comparison between our reconstruction of the temperature to other theoretical estimates of the temperature. 
We discuss these results and the possible systematics in Section \ref{sect:discussion}, before exposing our main conclusions.
\section{Simulation dataset}
\label{sect:Data}

\subsection{The NIKA2 LPSZ Twin Samples}
The cluster sample used in this work was selected within {\sc The Three Hundred} (The300 hereafter) suite of hydrodynamical simulations \citep{2018MNRAS.480.2898C}.
The300 simulations consists of zoom-in re-simulations of the 324 host regions of the most massive cluster-size halos from the dark-matter-only MultiDark Simulation run carried out with Planck cosmology \citep{2016A&A...594A..13P}.
Each cluster sits at the centre of boxes of $15 h^{-1}$ comoving Mpc of radius, and was simulated using runs of several different codes, at different resolutions.
We used the clusters from the runs performed using the {\sc Gadget-X} code, at low resolution ($3840^3$ dark matter particles).

The selection of the cluster sample used in this work, dubbed the NIKA2 LPSZ Twin Samples (NIKA2-TS, or NTS, hereafter), was chosen to match the statistical properties of the observed NIKA2 SZ Large Program (NIKA2 LPSZ) \citep{2020EPJWC.22800017M,2022EPJWC.25700038P}.
The NIKA2 LPSZ consists of a representative sample of 38 intermediate-redshift clusters ($z \in [0.5, 0.9]$) with masses spanning an order of magnitude from $10^{14}$ to $10^{15} \msun$, originally detected in the {\it Planck} and Atacama Cosmology Telescope (ACT) SZ cluster catalogues \citep{2016A&A...594A..27P,2013JCAP...07..008H}.
These clusters have been observed with the NIKA2 camera, in two bands of 150 GHz and 260 GHz. 
The angular resolution achieved by NIKA2 reaches $\sim17.6 "$ FWHM at 150 GHz, and $\sim 11.1 "$ FWHM at 260 GHz.
The synthetic maps used in this work are solely based on the maps at 150 GHz, as they are the ones maximizing the observed SZ signal from the clusters.

Selected within four snapshots between $z = 0.49$ and $z = 0.817$, to match the original NIKA2 LPSZ redshift coverage, the Twin Samples are three samples containing 38 clusters each. 
The three Twin Samples, denoted here NTS-$\mathrm{M_{tot}}$, NTS-$\mathrm{M_{HE}}$ and NTS-$\mathrm{Y_{500}}$ have been selected to match the NIKA2 LPSZ clusters' redshifts and their total mass $\mathrm{M_{tot, 500}}$, hydrostatic mass $\mathrm{M_{HE, 500}}$ and integrated Compton parameter $\mathrm{Y_{500}}$, respectively.
Some of the objects are in addition common between two or more NTS, leading to a final total of 96 simulated clusters.
The complete description of the selection and production of the Twin Samples can be found in \cite{2022EPJWC.25700036P}.

\subsection{Simulated maps}
This work required the use of different types of maps, which have been produced using The300 simulated data from the {\sc GadgetX}-3k runs. 
To produce the maps we selected only the gas particles counting as ICM, which could be responsible for X-ray and SZ signals. 
We thus selected the gas particles with a density below $10^{-1} \mathrm{cm}^{-3}$, below the star-formation density threshold, and with temperatures higher than 0.5 keV.
We focused on cubic regions centered on the main cluster of each simulation box, of $4\rth \times 4\rth \times 4\rth$ \footnote{The radius $\mathrm{R}_\Delta$ correspond to the radii within which the mean density corresponds to $\Delta$ times the critical density of the universe, $\rho_c = 3 H^2/8 \pi G$. Simirlarly, $\mathrm{M_{\Delta}}$ is the mass enclosed within $\mathrm{R_\Delta}$} in size.
We then produced theoretical maps of the SZ signal, X-ray signal (in truth the emission measure), temperature, using different codes for each type of map.
We also produced maps of the "effective length", necessary to convert the integrated squared electron density from the emission measure maps to a map of the integrated density.

The theoretical SZ maps, i.e. simulated maps of the Compton parameter $y_{tSZ}$ (defined in Eq. \ref{eq:y_def}), were produced using the {\tt pymsz} \citep{2012MNRAS.426..510C,2013MNRAS.430.3054C,2018MNRAS.480.2898C} python package. 
The code retrieves the thermodynamic quantities, including the density and temperature, to compute the pressure and in turn the theoretical Compton parameter $y_{tSZ}$ for all the selected particles. A Smoothed-Particle-Hydrodynamics (SPH) smoothing kernel is then applied to the Compton-$y_{tSZ}$ data.

The theoretical maps of the emission measure have been produced from the particles using the code {\tt Smac} \citep{2005MNRAS.363...29D,2020A&A...634A.113A}. The code smooths the field by applying an SPH kernel, then computes and projects the distribution of the squared gas density $\rho^2_{gas}$ along the line of sight, in units of $[\mathrm{gr^2cm^{-6}kpc^2cm}]$.
The output needs then to be converted to the standard units of the projected emission measure $[\mathrm{cm^{-6}Mpc}]$, using the mean molecular weight $\mu = 0.618$, the proton mass, and the electron-to-proton abundance ratio $n_e/n_p \sim 1.16$ (with slight variations in the field).

In order to assess the quality of the temperature maps reconstructed from the joint X-ray and SZ theoretical maps, we need to compare those to theoretical temperature maps.
We thus produced maps of the theoretical mass-weighted (see Eq. \ref{eq:tmw_def}) and spectroscopic-like (see Eqs. \ref{eq:tsl_def1} and \ref{eq:tsl_def2}) temperatures, starting from the temperatures, masses and densities of the particles. We run the SPH code {\sc Splash} \citep{2007PASA...24..159P} on the simulation files to obtain smoothed cubes, that we projected to generate maps.
The same code was used to produce maps of the effective length, defined below.

\section{Theoretical modelling}
\label{sect:theo_model}
\subsection{Reconstruction of the temperature}
The temperature measurement in a given sky direction resulting from the combination of SZ and X-ray data, denoted $\tszx$ hereafter, can be written as the ratio of a projected "effective pressure" $\effP$, obtained from SZ, and "effective density" $\effn$ obtained from X-ray :
\begin{equation}
    k_B \tszx = \frac{\effP}{\effn},
    \label{eq:tszx_def1}
\end{equation}
where $k_B$ is the Boltzmann constant.
These effective quantities link to the projected electronic pressure and electronic density as follows,
\begin{equation}
    \effP = \frac{1}{\leff} \int_{l.o.s.} P_e \diffd l
    \label{eq:effP_def}
\end{equation}
and
\begin{equation}
    \effn = \frac{1}{\leff} \int_{l.o.s.} n_e \diffd l,
    \label{eq:effn_def}
\end{equation}
through the effective length $\leff$, briefly discussed in Section \ref{sect:Data}.
Following the work of \citetalias{2017A&A...606A..64A}, the effective length is defined as 
\begin{equation}
    \leff = \frac{\left(\int_{l.o.s.} n_e \diffd l \right)^2}{\int_{l.o.s.} n_e^2 \diffd l}.
    \label{eq:leff_def}
\end{equation}
In a real observational setup, the maps of the projected pressure and squared density are obtained respectively from the SZ signal $y_{\mathrm{tSZ}}$ and X-ray surface brightness $S_X$ maps, with
\begin{equation}
    y_{\mathrm{tSZ}} = \frac{\sigma_T}{m_ec^2} \int_{l.o.s.} P_e \diffd l,
    \label{eq:y_def}
\end{equation}
 where $\sigma_T$ is the Thomson cross-section, $m_e$ the electron mass, $c$ the celerity of light and 
 \begin{equation}
    S_X = \frac{1}{4\pi(1+z)^4} \int_{l.o.s.} \Lambda(T_e,Z) n_e^2 \diffd l
    \label{eq:sx_def}
\end{equation}
with $\Lambda(T_e,Z)$ the cooling function depending on gas temperature $T_e$ and metallicity $Z$, taking into account the interstellar absorption and instrument response.
As a result, the complete expression for the reconstruction of $\tszx$ from Eq. \ref{eq:tszx_def1}, can be expanded following Eqs. \ref{eq:effP_def} to \ref{eq:sx_def} as :
\begin{equation}
    k_B \tszx =  \frac{\effP}{\effn} = \frac{m_ec^2}{\sigma_T} y_{\mathrm{tSZ}}\sqrt{\frac{\Lambda(T_e,Z)}{4\pi(1+z)^4 S_X \cdot \leff}}.
    \label{eq:tszx_obs_formula}
\end{equation}

However, as this work relies on a purely theoretical setup, with no observational contributions, the complete formula for $k_B \tszx$ differs slightly from Eq. \ref{eq:tszx_obs_formula}. 
Indeed, we obtain the theoretical effective density $\effn$ from maps of the emission measure, computed inside {\tt Smac} as :

\begin{equation}
    \mathrm{EM_\rho} = \sum_{i \in l.o.s.} \rho_{g,i}^2 V_i = \sum_{i \in l.o.s.} \rho_{g,i} m_{g,i},
\end{equation}
where $\rho_g$ is the gas mass density and $m_g$ the gas mass. We need to convert this map to a map of the projected squared electronic density, 
\begin{equation}
    \mathrm{EM}_{n_e} = \sum_{i \in l.o.s.} n_{e,i}^2 l_i.
\end{equation}
We use at this effect the relation $\rho_g = \mu m_p(n_e+n_p)$, so that the final maps we use are in units of $\mathrm{[cm^{-6}Mpc]}$. 

The resulting effective density thus writes :
\begin{equation}
    \effn = \sqrt{\frac{\mathrm{EM}_{n_e} \cdot 10^3}{\leff}},
\end{equation}
where the $10^3$ factor is a simply a conversion factor accounting for the emission measure being in $\mathrm{[cm^{-6}Mpc]}$, with $\leff$ being in kpc.
As a result, the expression for $k_B \tszx$ in the current setup is written as:

\begin{equation}
    \label{eq:tszx_sims_formula}
    k_B \tszx = \frac{\effP}{\effn} = \frac{m_ec^2}{\sigma_T} \frac{1}{\sqrt{\leff}} \frac{y_{tSZ}}{\sqrt{\mathrm{EM}_{n_e} \cdot 10^3}}
\end{equation}

As it appears in the description of our methods, we thus use three different kinds of maps $(y_{tSZ}, \mathrm{EM}_{n_e}, \leff)$, produced using three different codes.
These maps have of course different properties, with different pixel numbers, pixel sizes, and different angular resolutions.
In order to avoid inconsistencies between the maps due to these differences, we transform the $\mathrm{EM}_{n_e}$ and $\leff$ maps to match the properties of the $y_{tSZ}$ map. Indeed, as the latter possesses the lowest number of pixels, the largest angular pixel size, and the lowest angular resolution, we chose to degrade the other two maps to match the third.
Namely, we first resample the $\mathrm{EM}_{n_e}$ and $\leff$ maps so that the angular pixel size of both maps match that of the SZ map, i.e. with a pixel size of 3". 
We then match the spatial extension of the different maps by cropping the $\mathrm{EM}_{n_e}$ and $\leff$ maps to correspond to the extension of the SZ map of $16.65' \times 16.65'$.
Finally, we smooth all our input maps with a Gaussian kernel in order to match the angular resolution of the synthetic maps to the resolution of the NIKA2 camera at 150GHz.

To assess the quality of the temperature map inference using the joint X-ray and SZ maps, we need to compare it to other standard estimates of the temperature. 
We focus in this paper on the mass-weighted temperature $\tmw$, of which $\tszx$ should be the closest approximation \citepalias{2017A&A...606A..64A}, and on the spectroscopic-like temperature $\tsl$. 
Indeed, these two theoretical computations of the temperature should be the closest estimates of the total thermal energy in the ICM, and of an observed X-ray temperature, respectively \citep{2004MNRAS.354...10M, 2014MNRAS.439..588B,2018MNRAS.479.5385H,2023MNRAS.522..721P,2023A&A...675A.150Z}. 
It is however worth noting that depending on the simulation, the spectroscopic-like temperature may be biased with respect to the X-ray temperature obtained by fitting synthetic spectra, whether it is only for low temperatures or the full temperature range \citep{2014MNRAS.439..588B,2017MNRAS.470..166H,2019MNRAS.489.2439H,2023MNRAS.522..721P,2023A&A...675A.150Z}.

The mass-weighted temperature is defined as follows from \cite{2014MNRAS.439..588B} :
\begin{equation}
    \tmw = \frac{\sum_{i \in l.o.s.} T_i m_i \diffd l}{\sum_{i \in l.o.s.} m_i \diffd l},
    \label{eq:tmw_def}
\end{equation}
where $m$ is the mass of gas particles, 
while the spectroscopic-like temperature, from \cite{2004MNRAS.354...10M}, can be defined as :
\begin{equation}
    \tsl = \frac{\sum_{i \in l.o.s.} T_i w_i \diffd l}{\int_{i \in l.o.s.} w_i \diffd l}
    \label{eq:tsl_def1}
\end{equation}
where the weight $w_i$ is defined as
\begin{equation}
    w_i = \frac{n_{e,i}^2}{T_i^{3/4}}.
    \label{eq:tsl_def2}
\end{equation}

We produce the spectroscopic-like and mass-weighted temperature maps using {\sc Splash}, in a similar manner to the $\leff$ maps. 
In order to carry out a meaningful comparison, we in addition process the maps in the same way we process the input $(y_{tSZ}, \mathrm{EM}_{n_e}, \leff)$ maps, i.e. resizing, cropping and convolving them by a Gaussian kernel to match the properties of the $y_{tSZ}$ map.

\subsection{Modelling of the effective length}
\begin{table*}[ht]
    \centering
    \begin{tabular}{lccc}
    \hline \hline
    {\bf Parameter}  &  {\bf Median template} &  {\bf Relaxed template} &  {\bf Disturbed template}\\
    \hline
    $\mathbf{A_0}$  & $3.973_{-0.493}^{+1.122}$ & $4.012_{-0.346}^{+1.004}$ & $3.437_{-0.154}^{+0.438}$ \\
    $\mathbf{B_0}$  & $0.463\pm0.156$ & $0.534_{-0.168}^{+0.088}$ & $0.500_{-0.112}^{+0.058}$ \\
    $\boldsymbol{\alpha}$ & $0.501_{-0.031}^{+0.038}$ & $0.540_{-0.011}^{+0.013}$ & $0.447_{-0.009}^{+0.010}$ \\
    $\boldsymbol{\beta}$ & $5.531_{-2.542}^{+5.435}$ & $5.037_{-1.692}^{+3.599}$ & $8.508_{-3.305}^{+4.170}$ \\
    $\boldsymbol{\gamma}$ & $1.418_{-0.659}^{+1.920}$ & $1.083_{-0.293}^{+1.130}$ & $1.037_{-0.257}^{+0.981}$ \\
    \hline
    \end{tabular}
    \caption{Parameter values for the different GHP templates calibrated using the three combined Twin Samples, given with the 16th and 84th percentiles.}
    \label{tab:GHP_template_paramvals}
\end{table*}
\begin{figure}
    \centering
    \includegraphics[width=\linewidth, clip, trim = {0.75cm 0.25cm 0cm 0.75cm}]{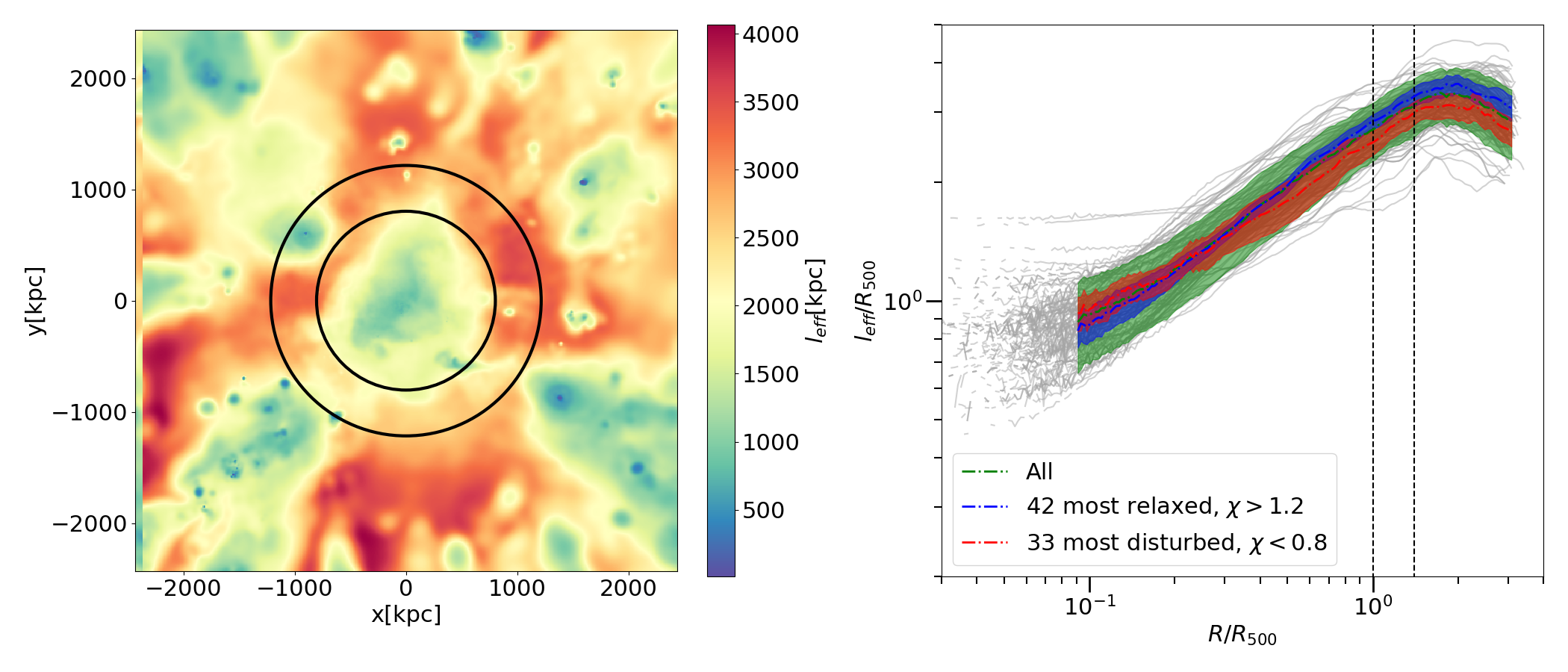}
    \caption{{\it Left}: Effective length map for the main cluster of region 211 of the snapshot 101, at $z = $0.817. The two rings correspond to $\rfh$ and $\rth$. {\it Right :} 1D profiles of the effective length for all clusters in our sample. We give in addition the median $\leff$ profiles computed for the full sample (green), the most relaxed (blue) and the most disturbed (red) clusters of the sample. The vertical dashed lines correspond to the $1.0 \rfh$ and $1.4 \rfh \sim \rth$.}
    \label{fig:leff_map_profiles}
\end{figure}

In the framework of hydrodynamical simulations, we have access to the true map of the effective length for each cluster, from which we can then reconstruct $\tszx$. 
However this quantity is not a direct observable, and as such, preparing an application of the reconstruction to real data requires first an effort to model $\leff$ to the required level of accuracy.

The most basic modelling that can be adopted for the effective length is simply assuming a constant value $l_0$ for the quantity, as was done as a test for instance in \citetalias{2017A&A...606A..64A} and Artis et al. (in prep).
Results from these two works suggest that this assumption may impact the normalisation of the temperature map (Artis et al. in prep) as well as the recovery of the substructures in the image \citepalias{2017A&A...606A..64A}, although the magnitude of this effect is not fully understood.
As a result, when trying to infer temperature maps with a good recovery of their spatial structure {\it and} accurate absolute measurements, a robust modelling of $\leff$ is crucial.

\citetalias{2017A&A...606A..64A} and Artis et al. (in prep) proposed a variety of different models in observational data, relying on the use of $\beta$ density profiles \citep{1976A&A....49..137C} fitted in the X-ray images.
In this work focused on the use of hydrodynamical simulations, we have access to "true" effective length maps, that we can use to infer and test a more accurate modelling of $\leff$.

For all the clusters in the sample, the effective length exhibits a similar behaviour, with low values of $\leff$ in the centre, and gradually increasing towards the outskirts, as we show in Fig.\ref{fig:leff_map_profiles}. 
All clusters also exhibit a flattening of the effective length, and in some cases a decrease, for radii $\gtrsim \rth$.
We also notice that relaxed clusters seem to have an effective length on average lower in the central regions than disturbed clusters, and higher in the outskirts.
This can be related to the shape of the density profiles, which are on average more peaked in the center for relaxed clusters \citep{Mostoghiu2019}.
In order to be able to fit this particular shape, we propose a functional form written as:

\begin{equation}
    \leff(R) = A_0 \rfh \frac{\left( B_0 \frac{R}{\rfh}\right)^\alpha}{\left(1+\left(B_0\frac{R}{\rfh}\right)^\beta\right)^{\gamma/\beta}}.
    \label{eq:GHP_def}
\end{equation}
The parameters $A_0$ and $\rfh/B_0$ represent respectively the amplitude and the cut-off radius of the $\leff$ profile, $\alpha$ the main power-law slope of the profile, and $\beta$ and $\gamma$ dictate the shape of the flattening and eventual decrease of the profile in the outskirts.
This functional form was derived as a "generalization" of the functional form for the transfer function of a first order high-pass filter, which exhibits the same features but restricted to the case $(\alpha = 1, \beta=2, \gamma=1)$, hence the name of "Generalized High Pass" (GHP) model that we will be using the rest of the paper for simplicity.
This functional form can also be found in the temperature profile proposed by \cite{2006ApJ...640..691V}.

An advantage of this approach is that once calibrated on simulations, this template profile, that can be propagated in a 2D template map, only requires the estimation of the $\rfh$ of a cluster to be applied directly in observations. 

We compute three different templates from this GHP model, calibrated on the full sample, on the most relaxed clusters, and the most disturbed clusters of the sample.
These will be respectively our Median, Relaxed and Disturbed templates.
We base our estimation of the cluster dynamical state on the relaxation parameter $\chi$ from \cite{2021MNRAS.504.5383D}, itself derived from the more general definition of \cite{2020MNRAS.492.6074H}. 
The definition of $\chi$ used in this paper is thus :

\begin{equation}
    \label{eq:dyn_state}
    \chi = \sqrt{\frac{2}{\left(\frac{\Delta_r}{0.1}\right)^2+\left(\frac{f_s}{0.1}\right)^2}},
\end{equation}
where $\Delta_r$ is the centre-of-mass offset and $f_s$ is the sub-halo mass-fraction. 
To segregate between the relaxed and disturbed clusters used to calibrate the templates, we use these values computed at $\rfh$, and selected as relaxed the clusters for which $\chi > 1.2$, and as disturbed those where $\chi < 0.8$.
Using this selection, the Relaxed template has been calibrated on 40 clusters, and the Disturbed template on 34 clusters, out of the 96 of the full sample (see Fig.\ref{fig:leff_map_profiles}).

In the end we give in Table \ref{tab:GHP_template_paramvals} the value of the fitted parameters for the three calibrated templates.
Some differences can be seen between the templates, notably on the amplitude $A_0$ and the sharpness of the "knee" $\beta$, although the different templates seem to agree on their general behaviours and properties.
\begin{figure*}[t!]
    \centering
    \includegraphics[width=0.9\linewidth, height = 8cm, trim = {10.5cm 3.5cm 10.5cm 5cm}, clip]{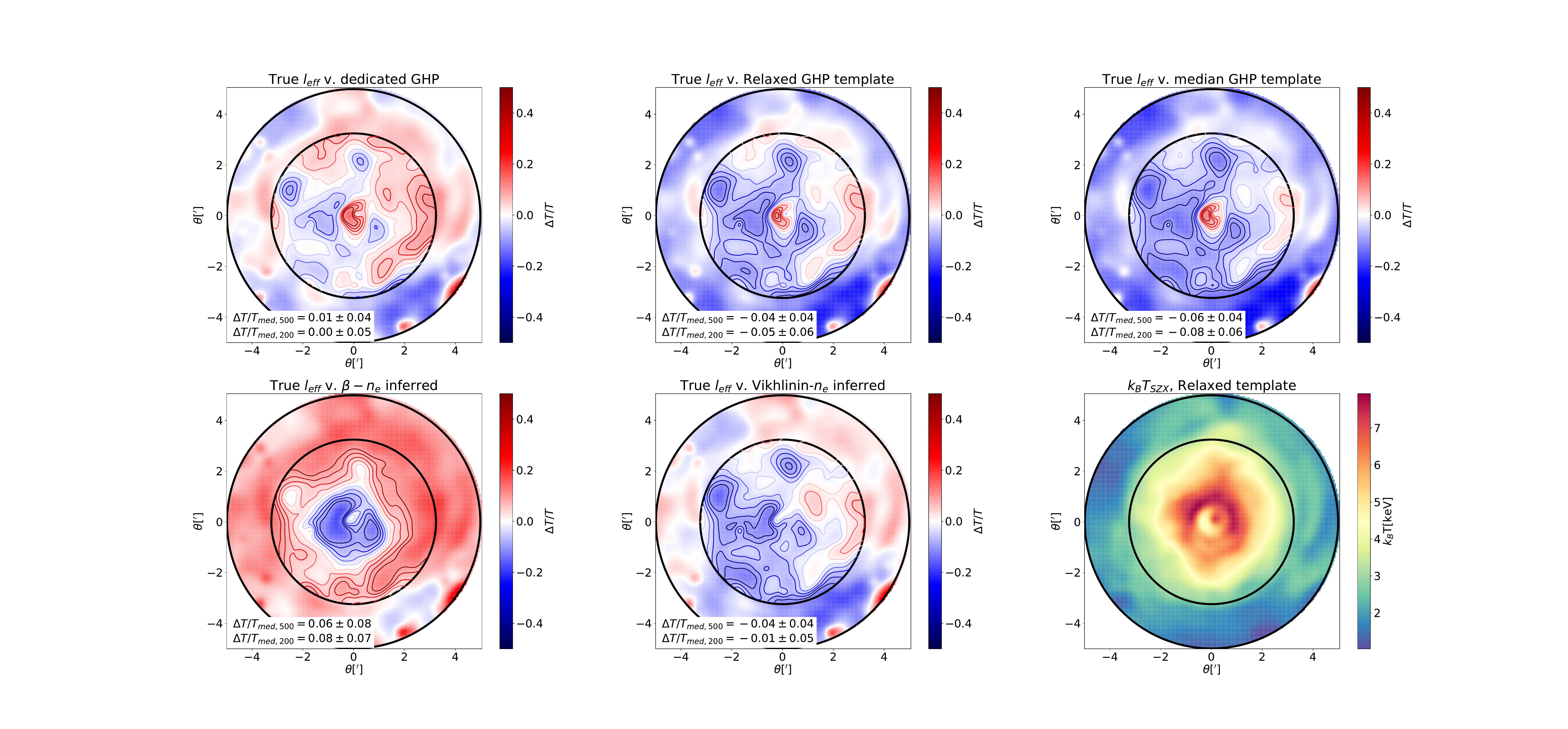}
    \includegraphics[width=0.9\linewidth, height = 8cm, trim = {10.5cm 3.5cm 10.5cm 5cm}, clip]{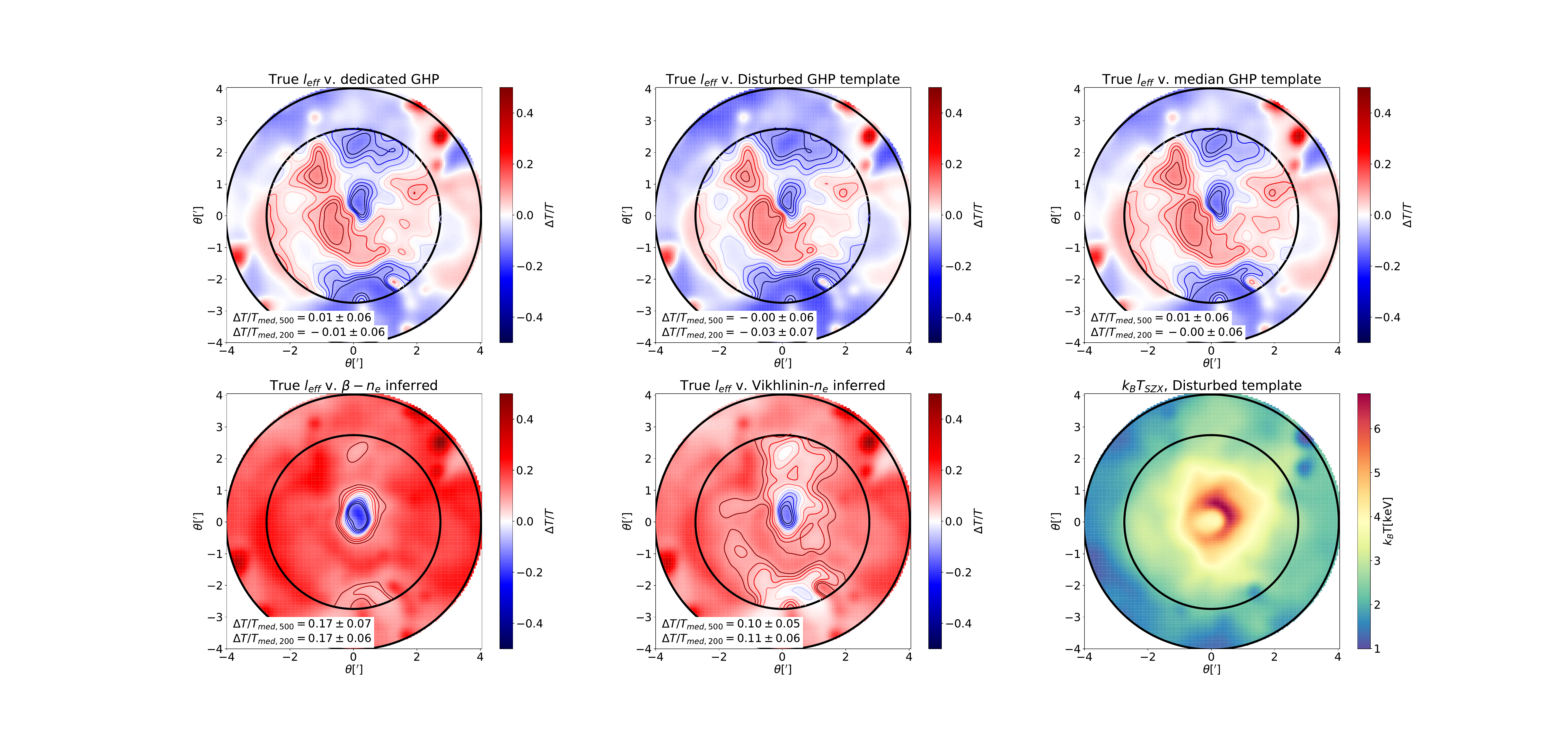}
    \caption{Two examples of comparison between $\tszx$ derived using the true $\leff$ and $\tszx$ obtained using a modelled effective length. The two rings correspond to $\rfh$ and $\rth$. The color scale of the comparative maps ranges from -50\% to +50\%. The levels of the contours inside $\rfh$ range from -10\% to +10\%.
    For each of the two clusters, the final panel shows the $\tszx$ map obtained using the optimal (Relaxed, Disturbed or Median) GHP template.\\
    {\it Top two rows:} Main cluster of the region 75 of the snapshot 110, at $z = $0.490. {\it Bottom two rows:} Main cluster of the region 135 of the snapshot 107, at $z = $0.592.}
    \label{fig:Tszx_templates_comparative_maps}
\end{figure*}

\subsection{Comparative performance between GHP and $n_e$-model inferred $\leff$}
\label{subsec:GHP_v_neinferred}
\begin{figure*}[ht]
    \centering
    \includegraphics[width=0.9\linewidth, height = 8.5cm, trim = {10cm 4cm 10.5cm 6cm}, clip]{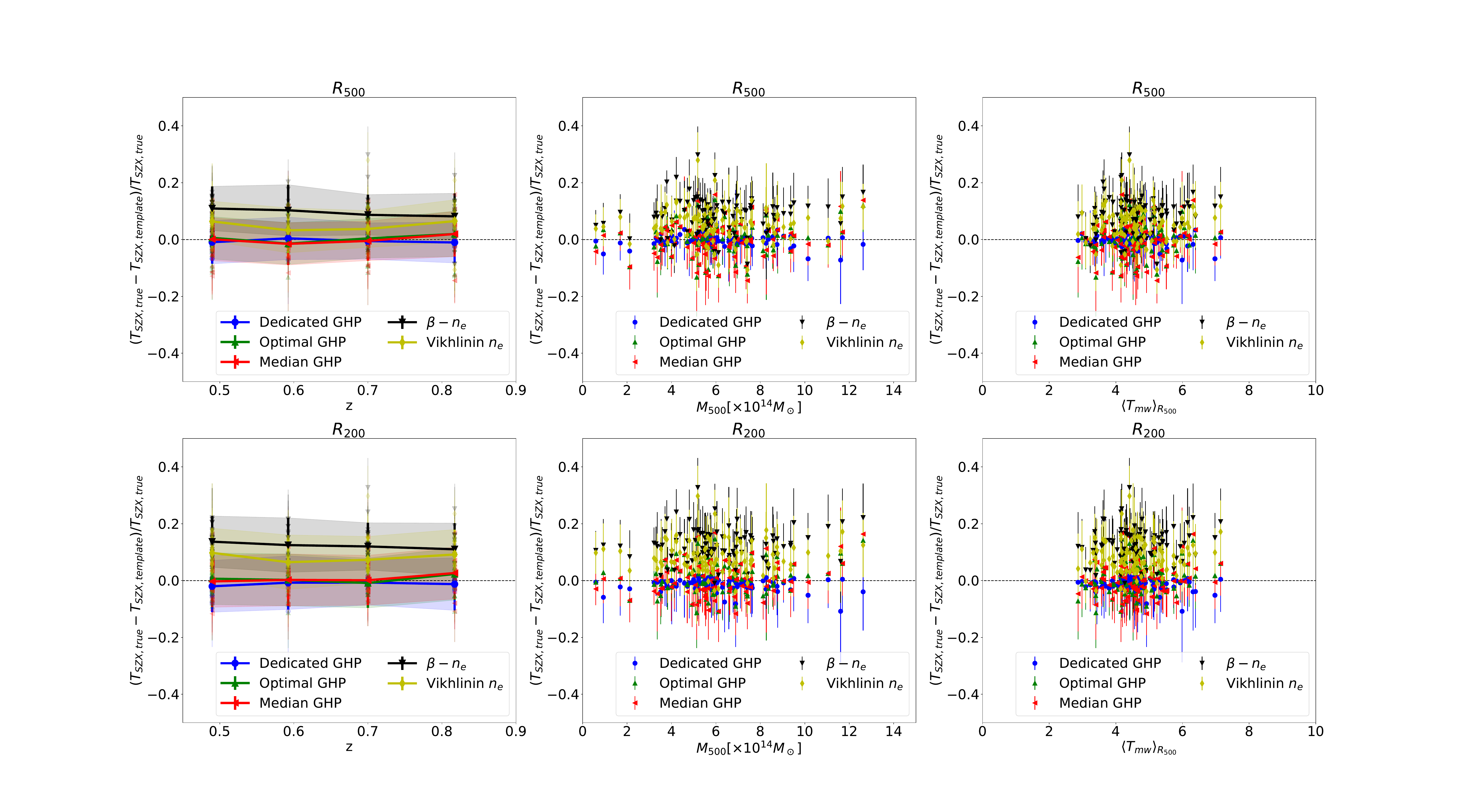}
    \caption{Median bias and scatter between $\mathrm{T_{SZX,true}}$ and $\mathrm{T_{SZX,template}}$, for all the clusters in the sample, depending on z (left column), $\mathrm{M_{500}}$ (middle column) and $\left<T_{mw}\right>_{R_{500}}$ (right column). 
    {\it Top row:} Biases and scatters taken at $\rfh$.
    {\it Bottom row:} Biases and scatters taken at $\rth$.}
    \label{fig:bias_stats_Tszx_Tszx}
\end{figure*}

We want to assess here the advantages of using a GHP template map calibrated on simulations with respect to the more standard approach based on the use of fitted density profiles.
To that end, we infer $\tszx$ temperature maps both using our GHP model and using 3D density profiles fitted using two different models, a $\beta$ model \citep[][given in Eq. \ref{eq:beta_model}]{1976A&A....49..137C} and a \cite{2006ApJ...640..691V} model (given in Eq. \ref{eq:vikhlinin_model}).
\begin{equation}
    \label{eq:beta_model}
    n_{e, \beta}(r) = n_0 \cdot \left(1+\left(\frac{r}{r_c}\right)^2\right)^{-3\beta/2}
\end{equation}

\begin{multline}
    \label{eq:vikhlinin_model}
    n_{e,Vikhlinin}(r) = \left[n_{0,1}^2 \left(\frac{r}{r_{c,1}}\right)^{-\alpha} \cdot \left(1+\left(\frac{r}{r_{c,1}}\right)^2\right)^{\frac{\alpha}{2}-3\beta_1} \cdot  \left(1+\left(\frac{r}{r_a}\right)^\gamma\right)^{-\frac{\epsilon}{2\gamma}}
\right.\\ 
    \left. + n_{0,2}^2\left(1+\left(\frac{r}{r_{c,2}}\right)^2\right)^{-3\beta_2} \right]^{1/2}
\end{multline}
We then compare the $\tszx$ maps reconstructed using parametric maps (i.e. using a GHP template or using density profiles) to the $\tszx$ map obtained using the true $\leff$ map, and estimate the bias and scatter around the "true" $\tszx$ resulting from the use of our models, with the bias $\Delta T/T$ defined as:
\begin{equation}
    \Delta T/T = \frac{T_{true} - T_{test}}{T_{true}},
\end{equation}
where $T_{true}$ is the temperature map obtained using the true effective length map, while $T_{test}$ is the temperature map obtained with different $\leff$ models.
In the end, the effective length models that we consider here are labeled as follows : 
"Dedicated GHP" corresponds to the $\leff$ map obtained by fitting directly the effective length profile for each cluster.
"Optimal GHP" corresponds to the best GHP template (Relaxed, Disturbed or Median) depending on the dynamical state of the cluster.
"$\beta-n_e$" corresponds to the effective length map obtained by fitting a 3D $\beta$ profile to the density.
"Vikhlinin $n_e$" corresponds to the effective length map obtained by fitting a 3D Vikhlinin profile to the density.
We show in Fig.\ref{fig:Tszx_templates_comparative_maps} the resulting maps for such a comparison in two particular cases (one relaxed cluster and a disturbed one), and in Fig.\ref{fig:bias_stats_Tszx_Tszx} the biases and scatters for all models of $\leff$, for all clusters. 
We show that the effective lengths obtained using GHP templates are statistically better suited to recover the "true" $\tszx$ that one would infer using the true effective length.
Indeed, even if the GHP reconstruction may be biased or very scattered around the true $\tszx$ for some particular clusters, and if the density profile $\leff$ can recover $\tszx$ with very low bias and scatter for some clusters (top two rows of Fig. \ref{fig:Tszx_templates_comparative_maps}), the GHP templates generally outperforms their $n_e$ profile counterparts, achieving a high accuracy, as reflected in the lower median bias, and the comparable scatter.
Indeed, whereas the median bias observed in the sample is below the percent level for the GHP templates, with a scatter of the order of 7\%, the temperatures inferred using a fitted density profile are biased by up to 10\%, with a comparable scatter.
It is worth noting that even if the $\leff$ model based on a Vikhlinin fit of the density profile manages a better recovery of $\tszx$ than when using a $\beta$ profile, it still underperforms with respect to the GHP templates.
In particular, the $\tszx$ values obtained using an effective length computed from fitted density profiles seem systematically underestimated. 
This would be the hint of an overestimation of $\leff$ within the chosen apertures of $\rfh$ and $\rth$.
When analysing the synthetic effective length maps for different models (see Fig. \ref{fig:GHP_v_Vik_leffmaps} in Appendix \ref{appendix:A}), we show that the synthetic $\leff$ maps based on density profiles seem to fail to drop to values as low in the center as the GHP templates.
In addition the maps obtained using parametric density profiles fail to reproduce the flattening and then decrease of the effective length, but rather keep on increasing beyond $\rth$. 

This effect, particularly that in the central regions, may be a signature of the gas clumping that fails to be well accounted for in the density profiles, but that we are able to capture when starting directly from the $\leff$ maps or profiles, to calibrate the GHP templates.
This explains to an extent the apparent overestimation of $\leff$, leading to an underestimated $\tszx$ when using the $\leff$ estimates based on density profiles.
It is in addition worth noting that this behaviour does not come with a dependence whether it is on the redshift, on the mass, or the mean temperature of the cluster, as shown in Fig.\ref{fig:bias_stats_Tszx_Tszx}.

We also want to stress here that the density profiles used to implement this approach are 3D density profiles computed in thin radial bins (we used 150 logarithmically spaced radial bins from 0.005 $\rth$ to $2\rth$).
In a more realistic, observational-like setup, where the density profiles are generally obtained using 2D deprojected data computed in fewer radial bins, this effect could be even stronger, highlighting the potentiality of using templates calibrated in simulations.

\section{Results}
\label{sect:results}

\begin{figure*}[ht]
    \centering
    \includegraphics[width=\linewidth, clip, trim={0 0 0 0}]{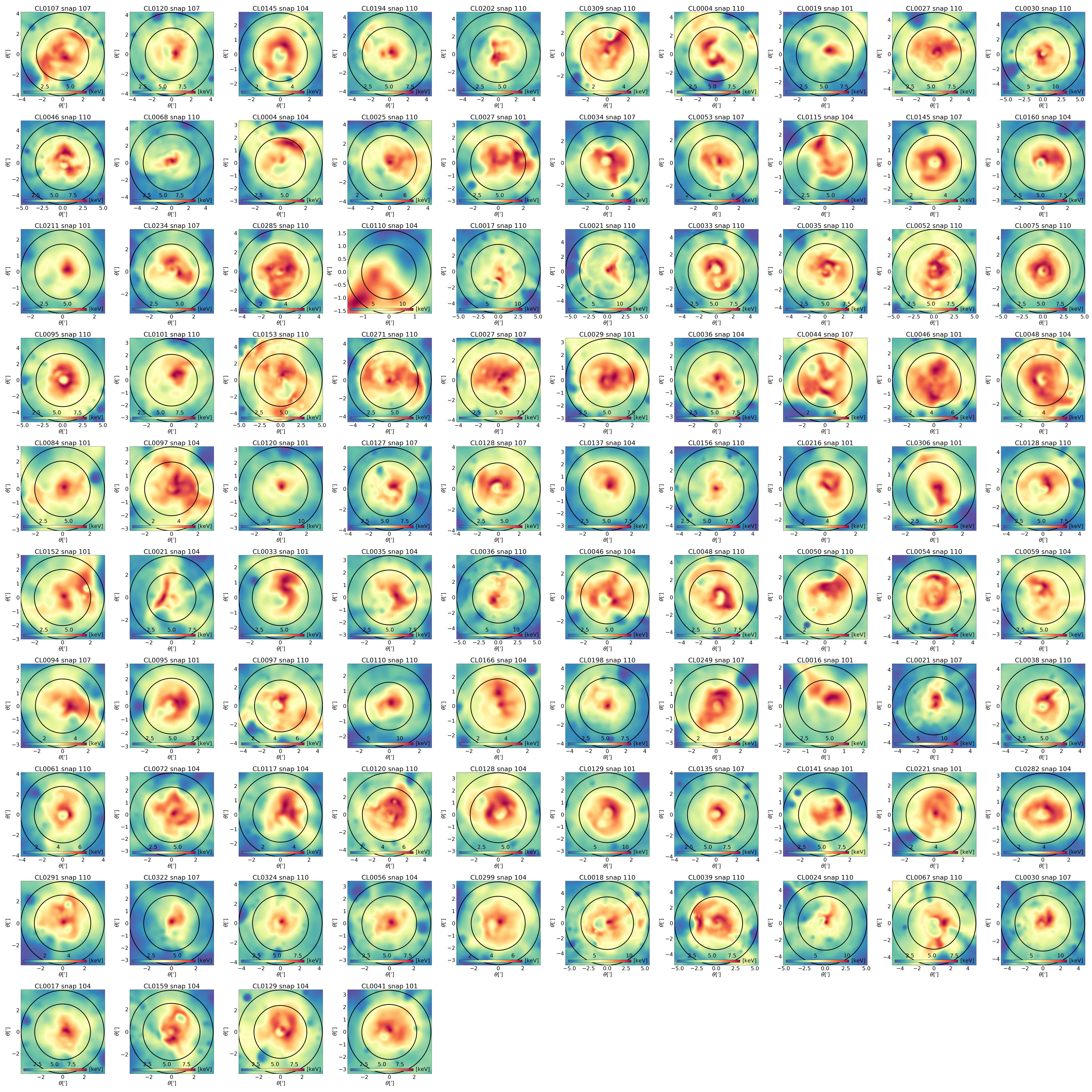}
    \caption{$\tszx$ maps obtained for all the clusters in our sample, using the optimal GHP template (Relaxed, Disturbed or Median) for each cluster.}
    \label{fig:all_tszx_maps}
\end{figure*}

\begin{figure*}
    \centering
    \includegraphics[width=0.9\linewidth, height = 8cm, clip, trim = {8cm 4.5cm 7cm 6cm}]{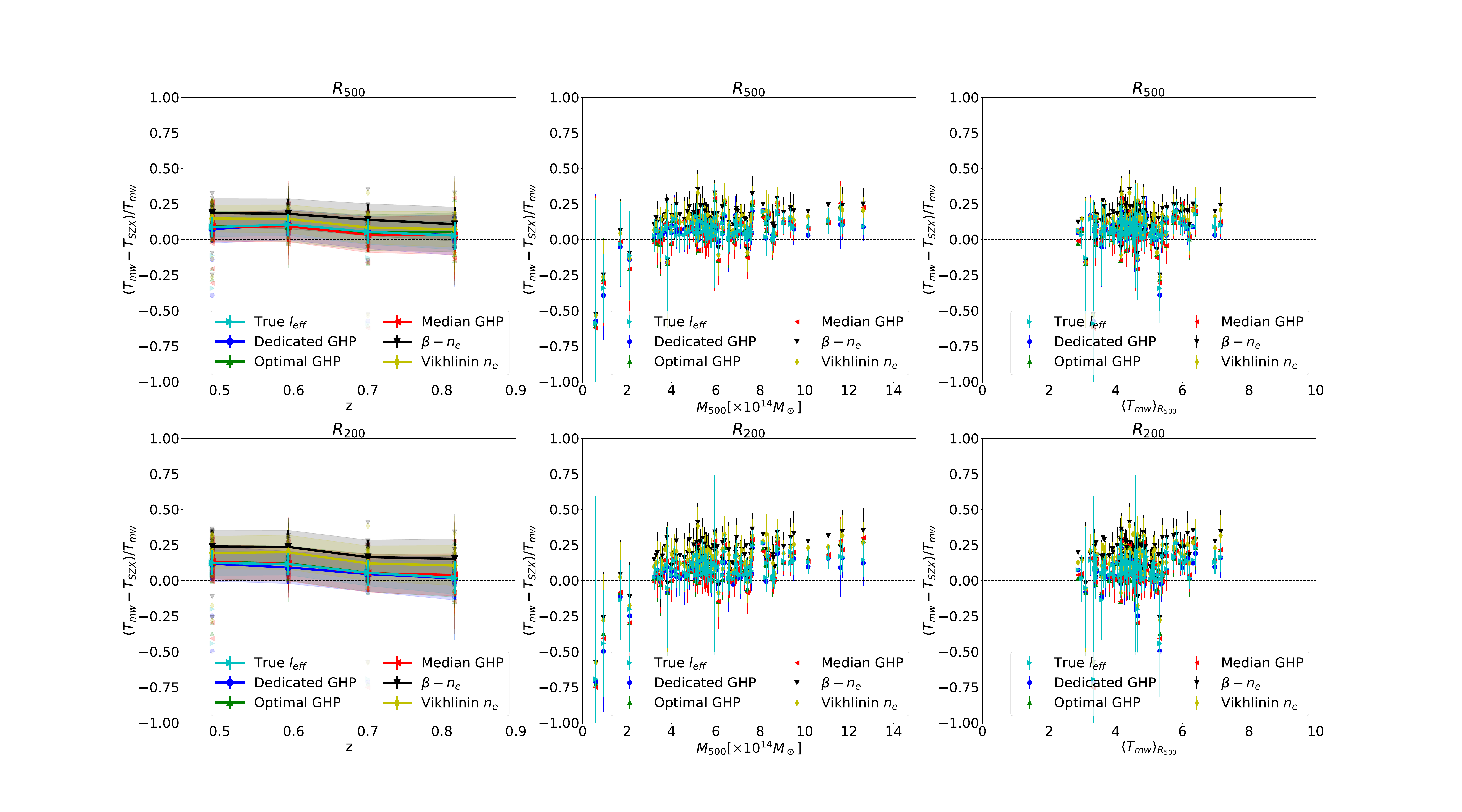}
    \caption{Comparison between $\tszx$, obtained using different models, and $\tmw$ for all the clusters in our sample, depending on their redshift, mass, and mean temperature at $\rfh$.
    {\it Top row :} Biases and scatters measured at $\rfh$. {\it Bottom row :} Biases and scatters measured at $\rth$. }
    \label{fig:bias_stats_Tszx_Tmw}
\end{figure*}

\begin{figure*}
    \centering
    \includegraphics[width=0.9\linewidth, height = 8cm, clip, trim = {8cm 4.5cm 7cm 6cm}]{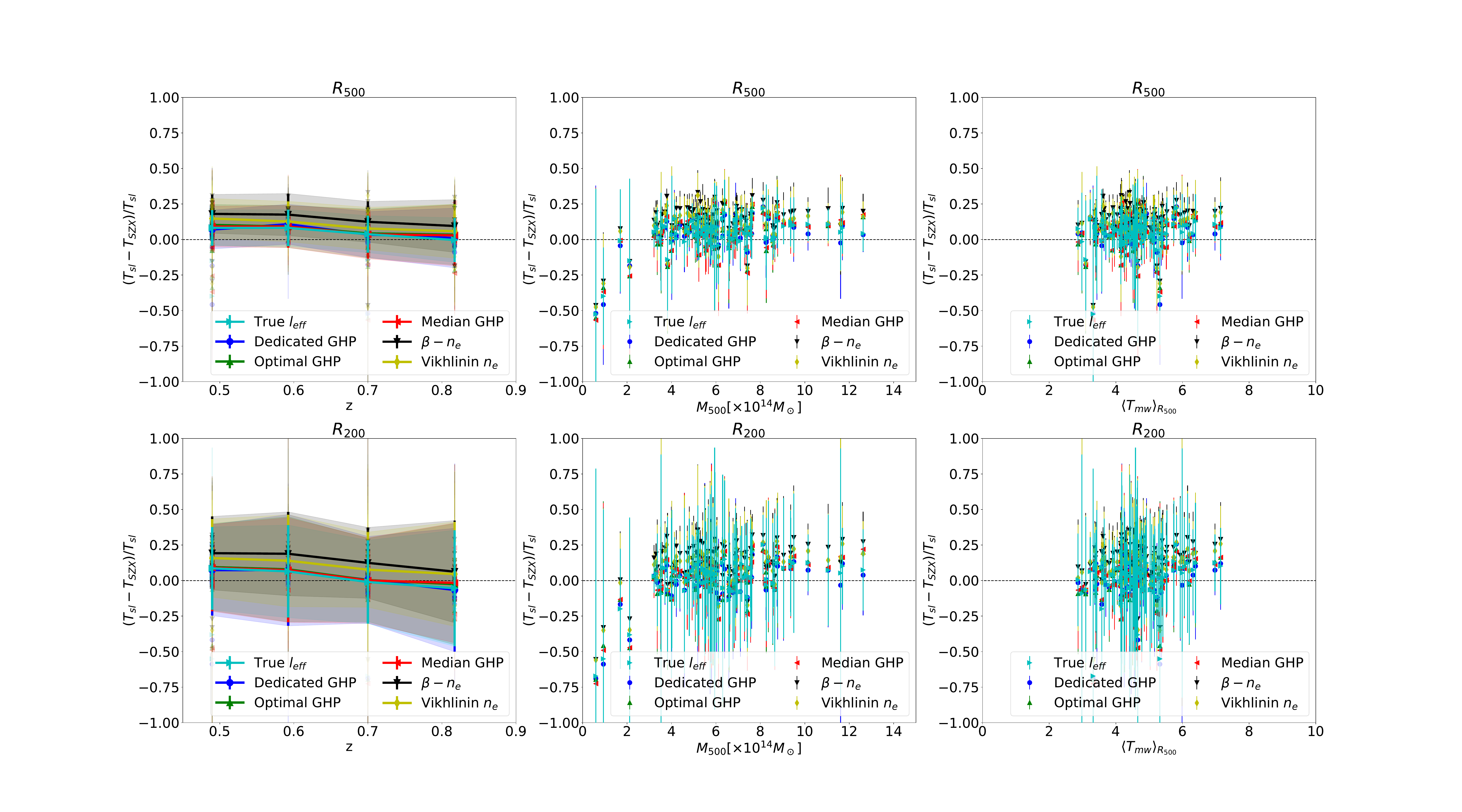}
    \caption{Comparison between $\tszx$, obtained using different models, and $\tsl$ for all the clusters in our sample, depending on their redshift, mass, and mean temperature at $\rfh$.
    {\it Top row :} Biases and scatters measured at $\rfh$. {\it Bottom row :} Biases and scatters measured at $\rth$. }
    \label{fig:bias_stats_Tszx_Tsl}
\end{figure*}
\begin{figure}[h]
    \centering
    \includegraphics[width=\linewidth, clip, trim = {4.5cm 2.75cm 4cm 3cm}]{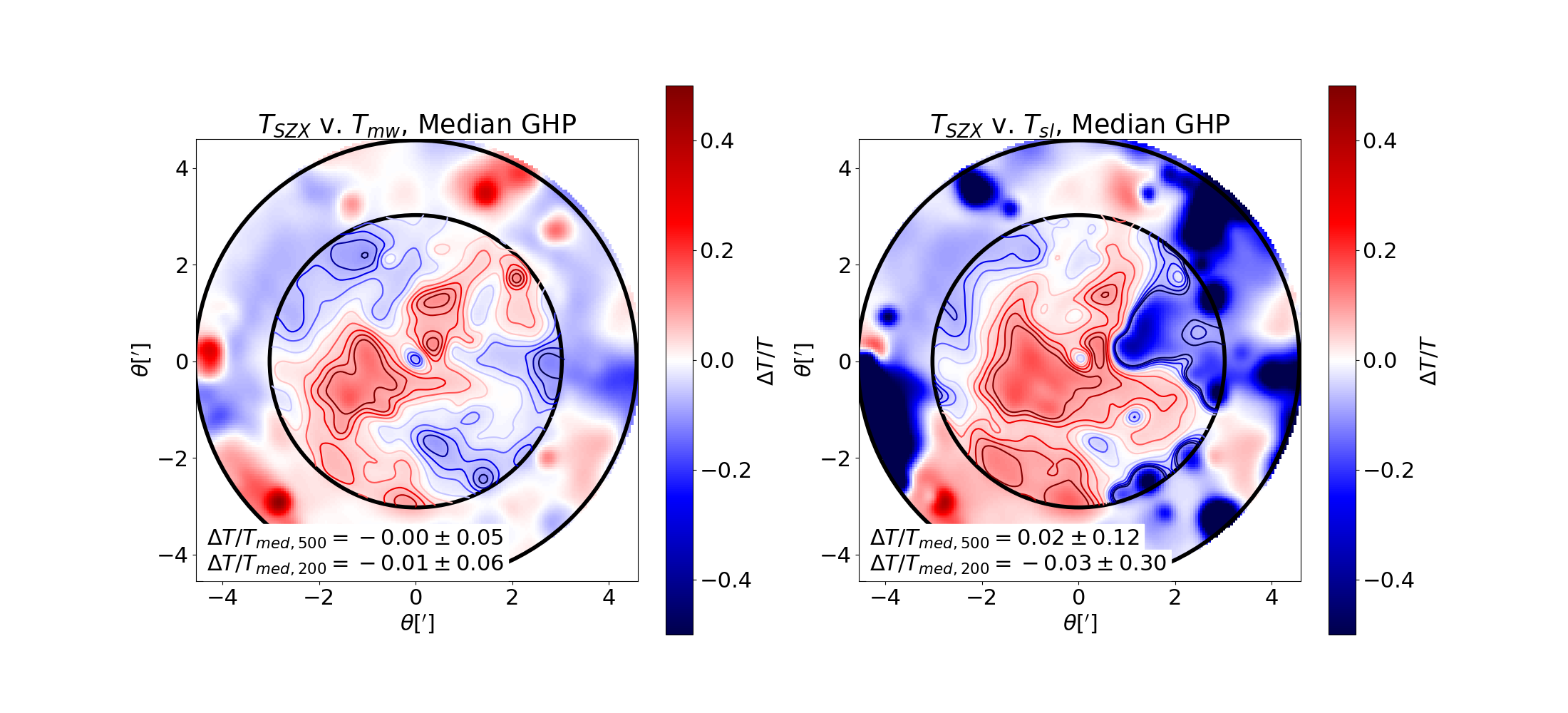}
    \caption{Comparison of $\tszx$ with the theoretical estimates of the temperature $\tmw$ (left) and $\tsl$ (right), for the main cluster of region 156 of the snap 110, at $z = $0.49. The color bar spans the $\pm 50\%$ range, the contours within $\rfh$ span the $\pm 10\%$ range.}
    \label{fig:tszx_v_ttheo}
\end{figure}
We give in Fig.\ref{fig:all_tszx_maps} the temperature maps for all the clusters in the combined Twin Samples. 
It appears that we are able to recover a comprehensive view of temperature maps for a great variety of cluster morphologies.
We notably show the presence of several obvious cool cores, that can be seen as paler cavities in the center of the concerned clusters.
We can also highlight overheated regions in several of those clusters.
Though a deeper analysis would be necessary to confirm with absolute certainty their nature, such regions may be related to shocks.
The potentiality of using the $\tszx$ approach to investigate the presence of shocks in galaxy clusters has incidentally been investigated by Artis et al. {\it in prep}.

We also desire to discuss the particular case of cluster CL0110 of the snapshot 104 (3rd row 4th column of Fig.\ref{fig:all_tszx_maps}). 
This particular cluster seems to exhibit a very peculiar temperature distribution, grossly mis-centered with respect to the alleged coordinates of the cluster center.
After careful investigation of this particular cluster, it appears that this effect is purely coming from the $\tszx$ derivation. 
Indeed, both the effectives quantities $(\effP, \effn)$ and the theoretical estimates of the temperature $(\tmw, \tsl)$ appear to have distributions well centered inside $\rfh$.
This cluster is a low mass system $(\mathrm{M_{500}} = 5.93 \times 10^{13} \mathrm{M_\odot})$ showing a quite peculiar and asymmetric distribution of its gas density, connected to a neighbour halo of similar size.
As a result of this complex architecture, the density distribution shows a non-negligible decrement in the bottom left corner of the map, artificially increasing the reconstructed $\tszx$. 
This examples highlights that in such peculiar systems, the reconstruction of $\tszx$ must be carried out with an extreme caution. 
It may also be the hint that in very low mass clusters, with lower gas densities, the $\tszx$ approach might not be so well suited to easily recover robust and reliable temperature estimates, as we develop later on.

In Figures \ref{fig:bias_stats_Tszx_Tmw} and \ref{fig:bias_stats_Tszx_Tsl} we compare $\tszx$ with the mass-weighted temperature $\tmw$ and the spectroscopic-like temperature $\tsl$, respectively. 
We show that there exists a slight offset between the theoretical estimates $(\tmw, \tsl)$ and our reconstruction of $\tszx$. 
Indeed, $\tszx$ seems to be slightly underestimated with respect to the theoretical estimates of the temperature, by $9.6 \pm 10.2\%$ at $z = $0.49 when using the optimal GHP template.
In addition, though this offset seems to be present both between $\tszx$ and $\tmw$ and $\tszx$ and $\tsl$ at comparable magnitude, the scatter of the relation is significantly higher between $\tszx$ and $\tsl$, especially when considering the full distribution within $\rth$. This can be also appreciated in comparative maps for single clusters, as we show in Fig.\ref{fig:tszx_v_ttheo}.
We however show a mild dependence of this offset with redshift, with temperature maps of clusters at high redshift seemingly less biased than those at lower redshifts, and fully compatible with no offset. 
Indeed, when considering the $\tszx$ map recovered using the optimal GHP template, the level of bias which is of $9.6 \pm 10.2\%$ at $z = $0.49, goes down to $5.1 \pm 13.8\%$ at $z = $0.817.
The most plausible explanation for this effect is the resolution of the maps at different redshifts.
Indeed, as the redshift goes up, the angular pixel size stays the same, meaning that the number of observed kpc per pixel is larger.
As a result, the number of pixels inside $\rfh$ and $\rth$ is lower at higher $z$, and the fraction of $\rfh$ and $\rth$ covered by the PSF is higher at higher $z$.
We discuss in depth this effect and source of systematics in Sect. \ref{subsec:resolution}.

Despite all these effects, we nevertheless show that our reconstruction of $\tszx$ always remains compatible with $\tmw$ and $\tsl$ within 1$\sigma$, especially in the case of the reconstruction using the GHP templates.
We also find no hint for dependencies on mass and temperature. 
We however show that for the four least massive clusters of the sample, and in particular the least massive cluster, CL0110 of snap 104, that we discussed in the previous paragraphs, the offset between $\tszx$ and the other estimates of the temperature is quite large, with a strong scatter.
This is another indication that although in our current mass range we see no evolution of the bias with mass, in the case of lower mass clusters, with lower densities and stronger sensitivity of the gas distribution to astrophysical effects, the reconstruction of $\tszx$ should be carried out with particular caution. 
In particular, it would seem that the capabilities of the $\tszx$ approach seem to start to break down for cluster masses $\lesssim \times 10^{14} \mathrm{M_\odot}$. This may be entirely driven by our choice of simulations, as the low resolution of the {\sc Gadget-X} 3k runs does not allow to fully resolve the physical processes in the smallest systems with masses $\lesssim \times 10^{13.5} \mathrm{M_\odot}$.
Though out of the scope of this precise paper, investigating the potentialities of the $\tszx$ approach in lower mass systems, in different simulations, may be needed to assess if a "range of validity" in mass exists for the $\tszx$ approach. 

\section{Discussion}
\label{sect:discussion}

\subsection{Calibration of the GHP templates}

In this work, we calibrate the GHP templates for $\leff$ on the three combined NIKA2 LPSZ Twin Samples. 
We have however {\it a priori} no reason to be sure that this calibration is valid for different samples.
As our end goal is to apply the $\tszx$ method in real observations, using the GHP templates calibrated in this paper, we need to test the adaptability of the templates to different samples.
To check for the sample dependency of our template calibration, we perform it separately on the three individual NTS.
The parameters for each of the fits are given in Table \ref{tab:complet_GHP_params}.

\begin{table*}[ht]
    \centering
    \begin{tabular}{lccccc}
    \hline \hline
     & $\boldsymbol{\mathrm{A_0}}$ & $\boldsymbol{\mathrm{B_0}}$ & $\boldsymbol{\alpha}$ & $\boldsymbol{\beta}$ & $\boldsymbol{\gamma}$ \\
     \hline
    {\bf Combined NTS} & & & & & \\
    \hline
    Relaxed Template & $4.012_{-0.346}^{+1.004}$ & $0.534_{-0.168}^{+0.088}$ & $0.540_{-0.011}^{+0.013}$ & $5.037_{-1.692}^{+3.599}$ & $1.083_{-0.293}^{+1.130}$ \\
    Median Template & $3.973_{-0.493}^{+1.122}$ & $0.463 \pm 0.156 $ & $0.501_{-0.031}^{+0.038}$ & $5.531_{-2.542}^{+5.435}$ & $1.418_{-0.659}^{+1.920}$ \\
    Disturbed Template & $3.437_{-0.154}^{+0.438}$ & $0.500_{-0.112}^{+0.058}$ & $0.447_{-0.009}^{+0.010}$ & $8.508_{-3.305}^{+4.170}$ & $1.037_{-0.257}^{+0.981}$ \\
     \hline
    {\bf $\boldsymbol{\mathrm{M_{tot}}}$ NTS} & & & & & \\
    \hline
    Relaxed Template & $5.168_{-1.075}^{+1.028}$ & $0.326_{-0.092}^{+0.176}$ & $0.529\pm0.009$ & $3.280_{-0.584}^{+1.263}$ & $2.062_{-1.134}^{+1.881}$ \\
    Median Template & $4.010_{-0.528}^{+1.189}$ & $0.452\pm 0.157$ & $0.507_{-0.028}^{+0.034}$ & $5.274_{-2.346}^{+5.471}$ & $1.400_{-0.661}^{+1.964}$ \\
    Disturbed Template & $3.838_{-0.339}^{+0.478}$ & $0.392_{-0.083}^{+0.088}$ & $0.460_{-0.009}^{+0.010}$ & $6.821_{-2.039}^{+4.091}$ & $1.642_{-0.784}^{+1.901}$ \\
     \hline
    {\bf $\boldsymbol{\mathrm{M_{HE}}}$ NTS} & & & & & \\
    \hline
    Relaxed Template & $3.660_{-0.110}^{+0.177}$ & $0.578_{-0.049}^{+0.039}$ & $0.544\pm0.008$ & $10.433_{-3.288}^{+3.055}$ & $0.825_{-0.111}^{+0.190}$ \\
    Median Template & $3.882_{-0.484}^{+1.060}$ & $0.475_{-0.159}^{+0.167}$ & $0.506_{-0.038}^{+0.044}$ & $5.715_{-2.628}^{+5.351}$ & $1.370_{-0.631}^{+1.934}$ \\
    Disturbed Template & $3.139_{-0.065}^{+0.068}$ & $0.659_{-0.033}^{+0.034}$ & $0.433\pm0.010$ & $12.768_{-2.535}^{+1.605}$ & $0.791_{-0.081}^{+0.093}$ \\
     \hline
    {\bf $\boldsymbol{\mathrm{Y_{500}}}$ NTS} & & & & & \\
    \hline
    Relaxed Template & $4.156_{-0.320}^{+0.656}$ & $0.551_{-0.111}^{+0.073}$ & $0.573_{-0.011}^{+0.013}$ & $5.028_{-1.301}^{+2.047}$ & $1.383_{-0.310}^{+0.758}$ \\
    Median Template & $4.060_{-0.582}^{+1.280}$ & $0.466_{-0.167}^{+0.179}$ & $0.506_{-0.027}^{+0.034}$ & $4.847_{-2.160}^{+5.388}$ & $1.498_{-0.721}^{+1.984}$  \\
    Disturbed Template & $6.674_{-1.399}^{+1.736}$ & $0.180_{-0.047}^{+0.089}$ & $0.495_{-0.023}^{+0.026}$ & $1.698_{-0.281}^{+0.403}$ & $3.211_{-1.229}^{+1.170}$ \\
    \hline
    \end{tabular}
    \caption{Parameter values for the GHP template calibrations in the different Twin Samples. The error bars correspond to the 16th and 84th percentiles.}
    \label{tab:complet_GHP_params}
\end{table*}

\begin{figure}
    \centering
    \includegraphics[width=\linewidth, clip, trim={4cm 3cm 4cm 3cm}]{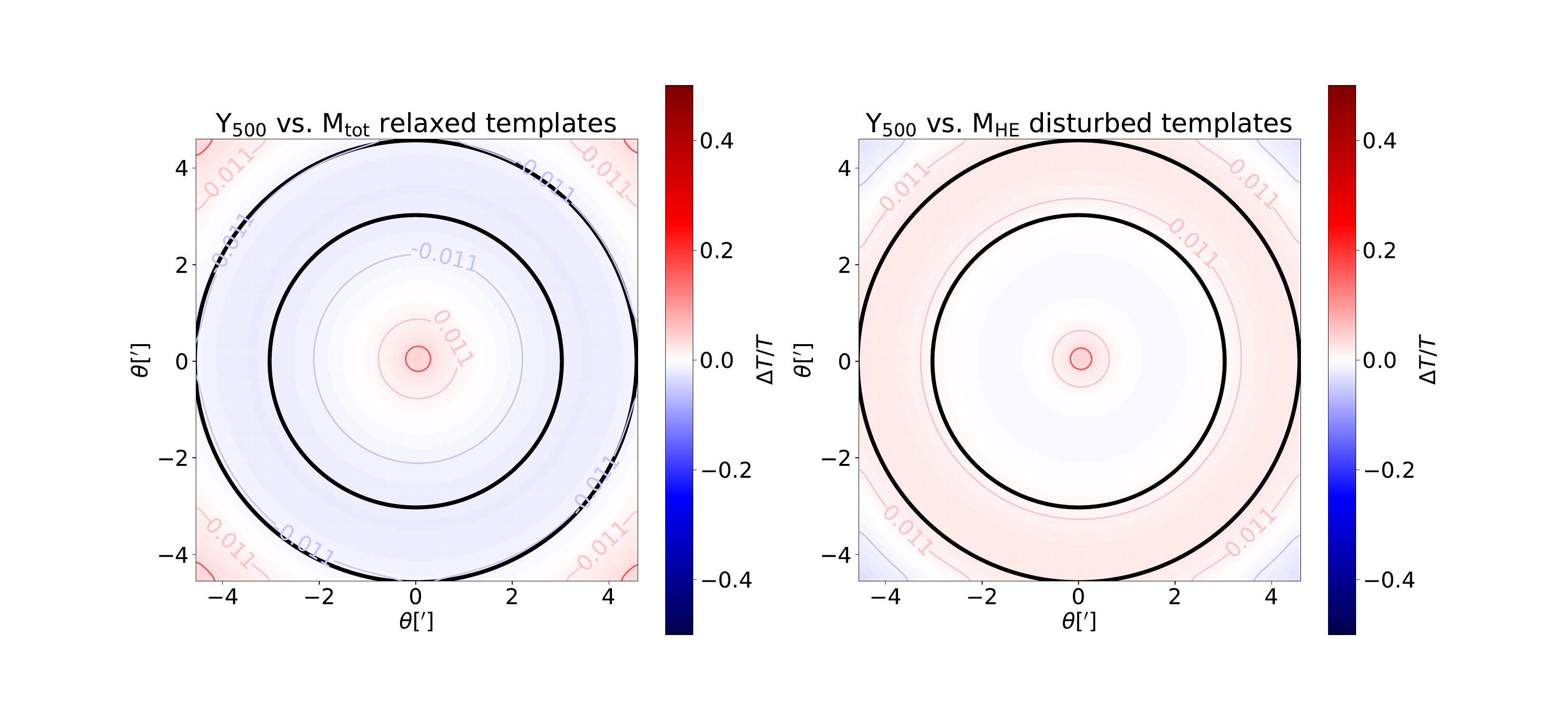}
    \caption{Comparison of the temperature maps obtained using GHP templates calibrated on different Twin Samples, for the main cluster of the region 156 of the snapshot 110, at $z = $0.49. We stress that the comparison showed here remains valid for all other clusters at all redshifts in our sample.
    {\it Left : } Comparison of the temperature maps obtained using the relaxed GHP templates calibrated on the $\mathrm{Y_{500}}$ and $\mathrm{M_{tot}}$ Twin Samples.
    {\it Right : } Comparison of the $\tszx$ maps obtained using the disturbed GHP templates calibrated on the $\mathrm{Y_{500}}$ and $\mathrm{M_{HE}}$ Twin Samples.}
    \label{fig:extreme_leff_templates}
\end{figure}

We show that the median profiles (i.e. calibrated on all clusters of the considered sample) of each NTS are all in perfect agreement with each other, and in agreement with the median template of the combined samples.
This agreement is however not so large between the relaxed and disturbed templates for the three different Twin Samples.
In particular, the relaxed templates of the $\mathrm{M_{tot}}$ and $\mathrm{Y_{500}}$ NTS and the disturbed templates of the $\mathrm{Y_{500}}$ and $\mathrm{M_{HE}}$ NTS show the largest discrepancies, with some parameters disagreeing at more than 2$\sigma$. 
These differences in the relaxed and disturbed templates, while the median templates are in perfect agreement, may stem from the selection of the Twin Samples, which has been performed solely on mass (or Compton parameter) and redshift, with no focus on the dynamical state.

This hints that the generalisation of the relaxed and disturbed templates to other samples should not be so trivial. 
This result needs to be accounted for when moving to the application on observed data.
We thus try to quantify the effect of these discrepancies on the final recovered temperature maps.
To that end, we compare the temperature maps obtained using the two extreme relaxed templates ($\mathrm{Y_{500}}$ and $\mathrm{M_{tot}}$) and disturbed templates ($\mathrm{Y_{500}}$ and $\mathrm{M_{HE}}$).
We show in Fig.\ref{fig:extreme_leff_templates} this comparison for one of the clusters of our sample.
As it could be expected, the relative difference between the temperature maps using the two extreme templates depends on the radius.
It appears that the maximum of the difference between the two maps is in the very center of the cluster, with a sharp increase in the most inner regions. 
However, the relative difference remains quite negligible, being always of the order of $\sim 1\%$, going up to slightly more than 3\% in the innermost regions. 
This effect is due to the degeneracies between some of the parameters of the GHP model, that allows to recover similar shapes of the $\leff$ profile for different sets of parameters.
This behaviour, that we show for one cluster, has been witnessed in all the clusters of the sample.
Compared to the median bias and scatter between $\tszx$ and $\tmw$ in our sample, which is of the order of 10\%, this contribution from the choice of calibration for the GHP templates thus appears to be negligible. 

\subsection{A word about the modelling of $\leff$}

\begin{figure}
    \centering
    \includegraphics[width=\linewidth, clip, trim = {1.25cm 1.25cm 1.25cm 1.15cm}]{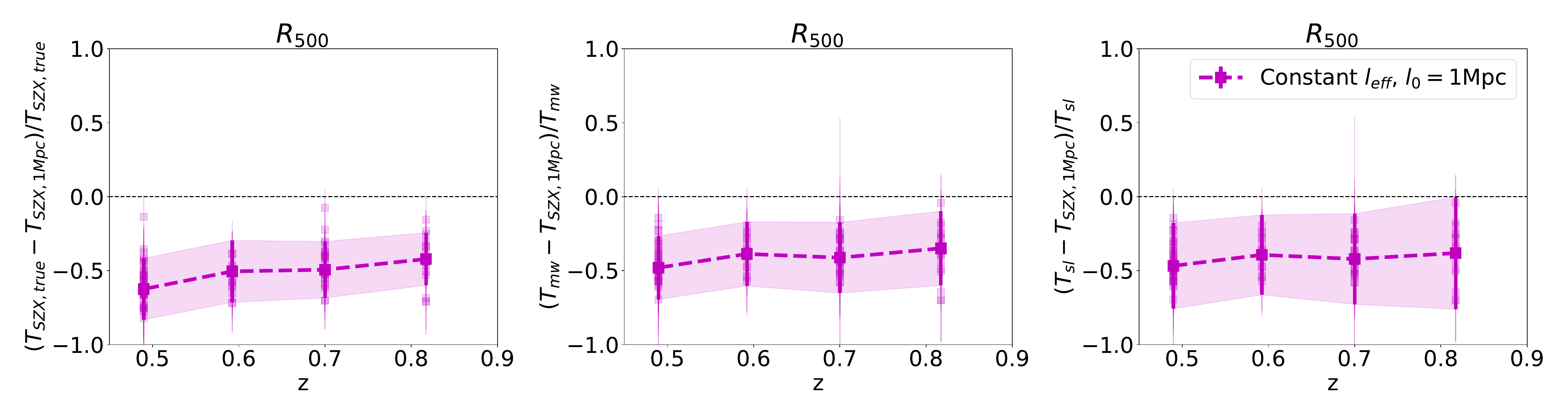}
    \caption{Comparison between $\tszx$ obtained using an effective length $l_0 = 1 \mathrm{Mpc}$ and $\tszx$ obtained using the true $\leff$ map (left), $\tmw$ (center) and $\tsl$ (right). 
    The biases and scatter are computed within $\rfh$.}
    \label{fig:1Mpc_leff}
\end{figure}

\begin{figure}
    \centering
    \includegraphics[width=\linewidth, clip, trim = {1cm 1.25cm 1cm 1.1cm}]{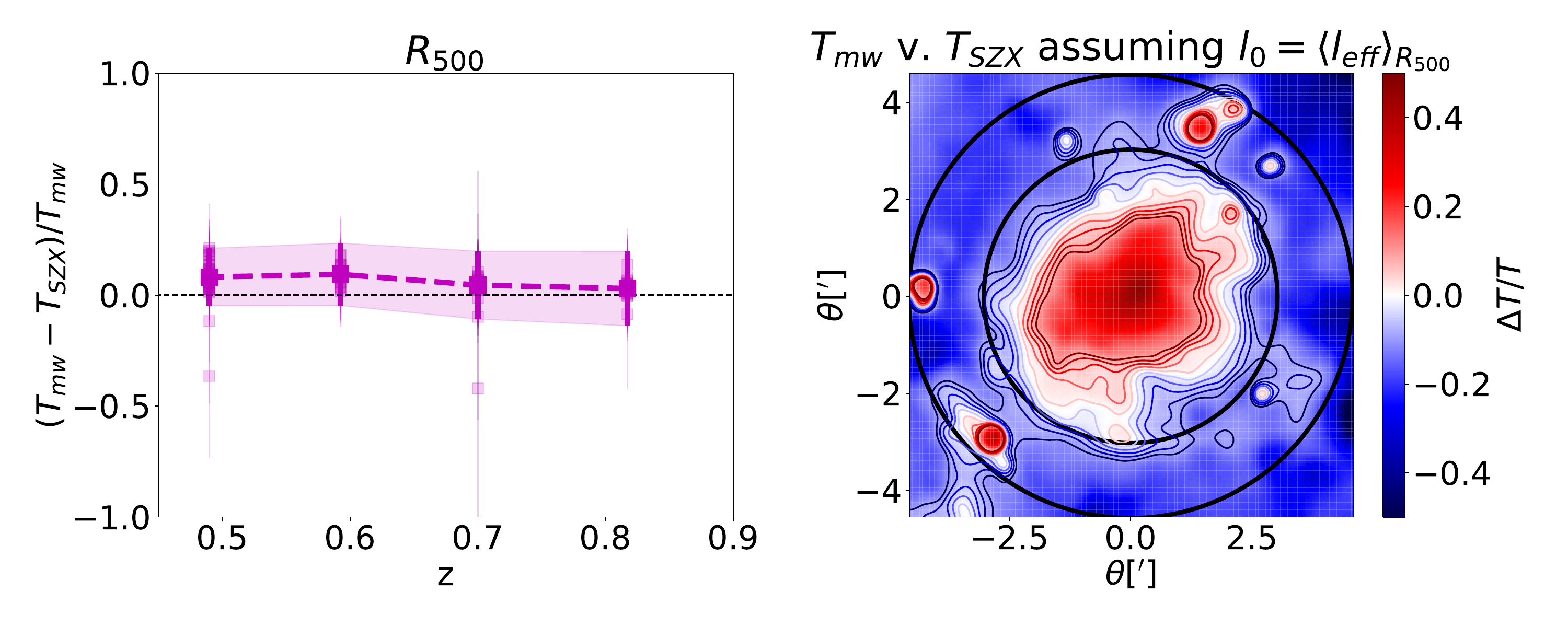}
    \caption{Comparison between $\tszx$ obtained using an effective length set to the mean of the true effective length at $\rfh$ and $\tmw$ for the full sample (left), and in the particular case of one cluster, the main cluster of region 156 of the snapshot 110 at $z = 0.49$ (right). 
    The biases and scatter are computed within $\rfh$.}
    \label{fig:mean_leff}
\end{figure}
Throughout this work, we assumed a variety of different models for the effective length, either relying on fitting a density profile, propagating it to a map of $\leff$, or using template maps calibrated on simulations, obtained with the GHP profile of $\leff$.
Past works \citepalias[][Artis et al. in prep]{2017A&A...606A..64A} however have sometimes used a more simple modelling for $\leff$, assuming the quantity to be a constant $l_0 \sim 1 \mathrm{Mpc}$ ($l_0 = 1.4 \mathrm{Mpc}$ in \citetalias{2017A&A...606A..64A}, based on the estimates of \cite{2013ApJ...778...52S} and \cite{2012ApJ...761...47M}).
We test here the effect of this assumption on the recovery of temperature maps.

We show in Fig.\ref{fig:1Mpc_leff} that assuming a constant $l_0 = 1 \mathrm{Mpc}$ leads to a bias of the order of $\sim 50\%$ with a high scatter of the order of 25\%.
One may of course argue that setting an effective length at 1Mpc when such a low value is barely reached in most of our $\leff$ maps is much too coarse of an approximation.
In order to use a more realistic modelling for a constant $\leff$, we test setting the value of $\leff$ at the mean value of the effective length within a given aperture (e.g. $\rfh$), instead of 1 $\mathrm{Mpc}$. 
Such a value is not a direct observable however it could be derived, for instance, from a density profile.
We show in Fig.\ref{fig:mean_leff} that in this case the bias between different reconstruction of temperatures is very low.
The scatter, however, is still quite high around $\sim 15\%$.
The most concerning issue with this modelling is that we show that the bias depends on the radius, with little of the global dynamics of the map being recovered properly (right panel of Fig.\ref{fig:mean_leff}).

This highlights that the simple modelling choice of assuming a constant effective length of $\sim 1 \mathrm{Mpc}$ should {\it not} be considered when trying to infer temperature maps, whether it is to study the structure or the normalization of the maps. 
It may however be enough to highlight overheated regions in maps, similarly to the approach of Artis et al. in prep.
Even when considering the mean value of the effective length, the goal of a study using this modelling should be clearly defined. 
If the purpose is to simply infer the normalization of a map, for instance to estimate integrated temperature values inside a given aperture, then assuming a constant effective length set at its mean value can be sufficient.
However when trying to study in addition the features and spatial distribution of the gas temperature, then a more complex modelling for the effective length {\it must} be considered, as assuming a constant $\leff$ will inevitably induce radially dependent biases in the final reconstruction.

\subsection{Impact of the resolution}
\label{subsec:resolution}
\begin{figure}
    \centering
    \includegraphics[width=0.8\linewidth, height = 7cm, clip, trim={0cm 1.75cm 2.5cm 3.25cm}]{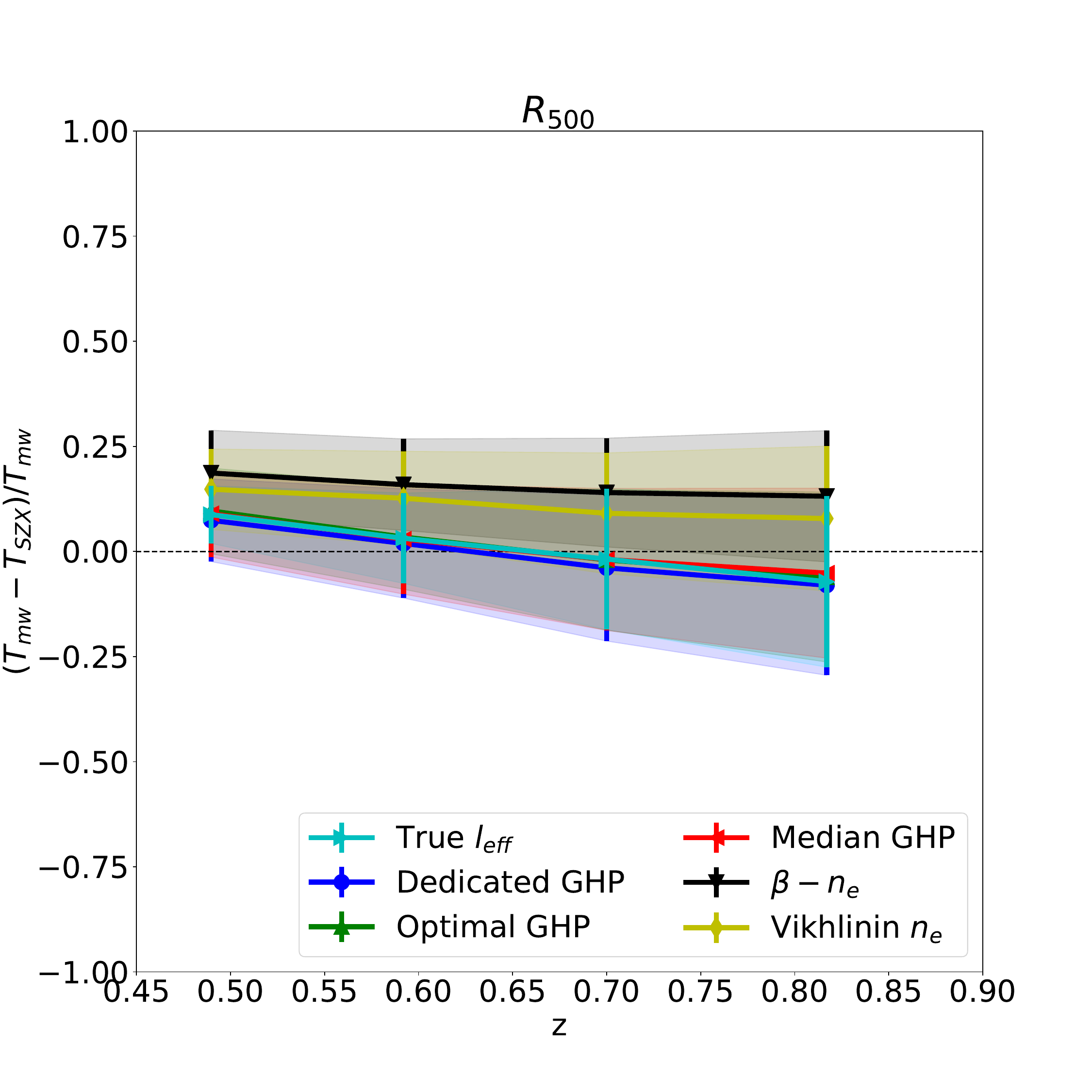}
    \caption{Comparison between $\tszx$ and $\tmw$ for the clusters of snapshot 110, displaced at redshifts $z = 0.49$ (original redshift), $z = 0.592$ (matching snapshot 107), $z = 0.7$ (matching snapshot 104) and $z = 0.817$ (matching snapshot 101).}
    \label{fig:snap_110_different_z}
\end{figure}

We mentioned in Sect. \ref{sect:results} the possible existence of a mild trend with redshift of the bias between $\tszx$ and the theoretical estimates of the temperature $\tmw$ and $\tsl$. 
We explore here the origin of this observed trend, and suggest that this behaviour is not a physical dependence on redshift, but rather purely due to the resolution and the sampling of the maps.
In our current setup, all maps have an angular pixel size of 3", and are smoothed with a Gaussian kernel of 18" FWHM.
For a cluster with $\rfh = 1000 \mathrm{kpc}$ at $z = 0.49$ (the lowest redshift of our simulated sample), $\sim 9020$ pixels will cover the area delimited by $\rfh$, and the area delimited by the FWHM of the PSF will cover 0.3\% of said area.
For the same cluster taken at $z = 0.817$ (the highest redshift available in our case), the area $\pi \rfh^2$ will be covered by $\sim 5770$ pixels and the area delimited by the FWHM of the PSF will cover 0.5\% of it.
Such a change is non negligible and may impact the final comparison between the reconstructed $\tszx$ and the theoretical $(\tmw, \tsl)$.

In order to investigate the impact of this effect, we chose to focus on the clusters at the lowest redshift of our sample, in snapshot 110, and pass them through our pipeline while displacing them at higher redshifts.
We stress here that we are always considering the same clusters and simply gradually reducing the sampling and resolution by artificially modifying their redshifts (thus modifying their angular diameter distance) in the pipeline, rather than actually look at the same haloes but taken at different steps of their evolution.
To remain comparable with the rest of the study, we chose to set the clusters at the same four redshifts than in the main analysis, i.e. $z = 0.49$ (snapshot 110, original redshift of the clusters), $z = 0.592$ (snapshot 107), $z = 0.7$ (snapshot 104) and $z = 0.817$ (snapshot 101, highest redshift available in the sample).

We show that by taking the same clusters and gradually lowering their sampling and resolution by mimicking the placement at different redshifts, we are indeed able to replicate a negative trend of the bias between $\tszx$ and $\tmw$ with redshift.
The $\tszx$ maps obtained using density profiles seem to be slightly less sensitive to this effect than the $\tszx$ maps obtained using GHP templates or using the true $\leff$ map.
As the former $\leff$ maps are "simpler" than the latter ones, with fewer distinctive features, it is possible that the gradual degradation of the maps induces a slightly lower loss of information.

We thus claim that the slight redshift trend with $z$ of the bias between $\tszx$ and $(\tmw, \tsl)$ that we highlighted in Sect. \ref{sect:results} is not a physical effect, but rather mainly triggered by the different sampling and resolution at different redshifts. 

\section{Conclusions}
Mapping of galaxy cluster gas temperature, although critical for the characterisation of the ICM and cluster mass estimates, remain a challenge. 
Few previous works have highlighted the potentiality of using a combination of X-ray and SZ data to derive temperature profiles, and fewer still tried to apply this method to map the temperature in 2D. However this approach focused on 2D maps had yet to be validated and tested on a large sample. 
In this work, we prepare the systematic application of this method to real data, by validating the theoretical aspects of the $\tszx$ reconstruction inside The300 suite of hydrodynamical simulations.
We start from maps of the theoretical $y_{tSZ}$ signal and of the gas emission measure, serving as proxies for the pressure and squared density maps, respectively.

We in addition have to take into account the cluster effective length, allowing to convert the map of the integrated squared density obtained from the emission measure to a map of the integrated density.
As this quantity is not a direct observable, we proposed a model, dubbed "Generalized High Pass" (GHP).
We calibrated our model on simulated maps of the true effective length, deriving three different template parametrizations calibrated on relaxed clusters, on disturbed ones, and on all clusters, to be applied as a median template.
We compared the $\tszx$ obtained using our modelling to other $\tszx$ maps obtained using more standard modellings, i.e. assuming either a constant value or parametric density profiles to infer the effective length.
We show that using the GHP model of $\leff$ for the reconstruction of $\tszx$ systematically outperforms other more standard modellings of the effective length, with a lower bias and scatter around the mean value of the temperature.
We in particular highlight how modelling the effective length with a single constant value should only be done with extreme caution with respect to the goals of the analysis, if at all. 
Indeed we show that carelessly choosing a constant value for $\leff$ may lead to extremely high biases and scatter in the recovered temperature, and that even with a carefully chosen constant value, a non-negligible radial bias remained in the $\tszx$ reconstruction.

Comparing our $\tszx$ reconstructions to theoretical estimates of the temperature, we find a good agreement with the mass-weighted and spectroscopic-like temperatures $\tmw$ and $\tsl$.
We find a median bias at most of the order of 10\% when using GHP templates, with a scatter of the same order of magnitude, with no hint of evolution with the mass or the mean gas temperature.
We also highlight a slight hint for a redshift dependence of the bias between $\tszx$ and the theoretical temperatures.
This is however fully driven by the sampling of the maps at different redshifts, with higher redshift clusters having a lower number of pixels within $\rfh$ and the PSF covering a larger fraction of the radius. 

This present work is the first of two papers of which the global aim is validating the $\tszx$ approach and applying it to high resolution tSZ maps, the NIKA2 LPSZ sample.
The next paper will be focused on the application to said data, using our own modelling of the effective length, and will discuss the impact of the instrumental effects on the final temperature maps.
The effects that will be discussed in particular will be that of the noise and the SZ transfer function.
Indeed, due to the scanning strategy of the SZ data, the large scales will be filtered out.
This needs to be properly modelled and understood to avoid introducing biases, notably radially dependent ones, in the final temperature maps.
Based on the recovered temperature maps we will investigate the properties of the gas distribution, including highlighting potential shocks and cool cores. 
Interesting should be the comparison of our approach on one LPSZ cluster, PSZ2G091.83+26.11, already analysed in the work of Artis et al.
They derived the temperature map for one of the LPSZ clusters, using a more simple modelling for the effective length, and used it to highlight an overheated region of the cluster, trying to link it to a shock.

\begin{acknowledgements}
 The authors acknowledge membership to {\sc The Three Hundred} collaboration, thanks to which this work was made possible. The simulations used in this paper have been performed in the MareNostrum Supercomputer at the Barcelona Supercomputing Center, thanks to CPU time granted by the Red Española de Supercomputación. As part of {\sc The Three Hundred} project, this work has received financial support from the European Union’s Horizon 2020 Research and Innovation programme under the Marie Skłodowska-Curie grant agreement number 734374, the LACEGAL project.  
 R.W., M.D.P. and A.F. acknowledge financial support from PRIN-MUR grant 20228B938N {\it"Mass and selection biases of galaxy clusters: a multi-probe approach"} funded by the European Union Next generation EU, Mission 4 Component 1  CUP C53D2300092 0006.
 R. Adam acknowledges support from the Programme National Cosmologie et Galaxies (PNCG) of CNRS/INSU with INP and IN2P3, co-funded by CEA and CNES. 
 R. Adam was supported by the French government through the France 2030 investment plan managed by the National Research Agency (ANR), as part of the Initiative of Excellence of Université Côte d'Azur under reference number ANR-15-IDEX-01.
 GY and WC  would like to thank Ministerio de Ciencia e Innovación (MICINN) for financial support under the project grant PID2021-122603NB-C21. WC is further supported by Atracci\'{o}n de Talento fellowship no. 2020-T1/TIC19882 granted by the Comunidad de Madrid and by the Consolidación Investigadora no. CNS2024-154838 granted by the Agencia Estatal de Investigación (AEI) in Spain, ERC: HORIZON-TMA-MSCA-SE for supporting the LACEGAL-III (Latin American Chinese European Galaxy Formation Network) project with grant number 101086388.
 M.M.E. acknowledges the support of the French Agence Nationale de la Recherche (ANR), under grant ANR-22-CE31-0010.
\end{acknowledgements}

%
%

\bibliographystyle{aa}
\bibliography{theo_tszx_biblio}

@ARTICLE{2016ApJ...827..112B,
       author = {{Biffi}, V. and {Borgani}, S. and {Murante}, G. and {Rasia}, E. and {Planelles}, S. and {Granato}, G.~L. and {Ragone-Figueroa}, C. and {Beck}, A.~M. and {Gaspari}, M. and {Dolag}, K.},
        title = "{On the Nature of Hydrostatic Equilibrium in Galaxy Clusters}",
      journal = {The Astrophysical Journal},
         year = 2016,
        month = aug,
       volume = {827},
       number = {2},
          eid = {112},
        pages = {112},
          doi = {10.3847/0004-637X/827/2/112},
archivePrefix = {arXiv},
       eprint = {1606.02293},
 primaryClass = {astro-ph.CO},
       adsurl = {https://ui.adsabs.harvard.edu/abs/2016ApJ...827..112B}
}

@ARTICLE{1993Natur.366..429W,
       author = {{White}, Simon D.~M. and {Navarro}, Julio F. and {Evrard}, August E. and {Frenk}, Carlos S.},
        title = "{The baryon content of galaxy clusters: a challenge to cosmological orthodoxy}",
      journal = {Nature},
         year = 1993,
        month = dec,
       volume = {366},
       number = {6454},
        pages = {429-433},
          doi = {10.1038/366429a0},
       adsurl = {https://ui.adsabs.harvard.edu/abs/1993Natur.366..429W}
}

@ARTICLE{2019SSRv..215...25P,
       author = {{Pratt}, G.~W. and {Arnaud}, M. and {Biviano}, A. and {Eckert}, D. and {Ettori}, S. and {Nagai}, D. and {Okabe}, N. and {Reiprich}, T.~H.},
        title = "{The Galaxy Cluster Mass Scale and Its Impact on Cosmological Constraints from the Cluster Population}",
      journal = {Space Science Reviews},
         year = 2019,
        month = feb,
       volume = {215},
       number = {2},
          eid = {25},
        pages = {25},
          doi = {10.1007/s11214-019-0591-0},
archivePrefix = {arXiv},
       eprint = {1902.10837},
 primaryClass = {astro-ph.CO},
       adsurl = {https://ui.adsabs.harvard.edu/abs/2019SSRv..215...25P}
}

@ARTICLE{2011ARA&A..49..409A,
       author = {{Allen}, Steven W. and {Evrard}, August E. and {Mantz}, Adam B.},
        title = "{Cosmological Parameters from Observations of Galaxy Clusters}",
      journal = {Annual Review of Astronomy and Astrophysics},
         year = 2011,
        month = sep,
       volume = {49},
       number = {1},
        pages = {409-470},
          doi = {10.1146/annurev-astro-081710-102514},
archivePrefix = {arXiv},
       eprint = {1103.4829},
 primaryClass = {astro-ph.CO},
       adsurl = {https://ui.adsabs.harvard.edu/abs/2011ARA&A..49..409A}
}

@ARTICLE{2006MNRAS.369.2013R,
       author = {{Rasia}, E. and {Ettori}, S. and {Moscardini}, L. and {Mazzotta}, P. and {Borgani}, S. and {Dolag}, K. and {Tormen}, G. and {Cheng}, L.~M. and {Diaferio}, A.},
        title = "{Systematics in the X-ray cluster mass estimators}",
      journal = {Monthly Notices of the Royal Astronomical Society},
         year = 2006,
        month = jul,
       volume = {369},
       number = {4},
        pages = {2013-2024},
          doi = {10.1111/j.1365-2966.2006.10466.x},
archivePrefix = {arXiv},
       eprint = {astro-ph/0602434},
 primaryClass = {astro-ph},
       adsurl = {https://ui.adsabs.harvard.edu/abs/2006MNRAS.369.2013R}
}

@ARTICLE{2009ApJ...705.1129L,
       author = {{Lau}, Erwin T. and {Kravtsov}, Andrey V. and {Nagai}, Daisuke},
        title = "{Residual Gas Motions in the Intracluster Medium and Bias in Hydrostatic Measurements of Mass Profiles of Clusters}",
      journal = {The Astrophysical Journal},
         year = 2009,
        month = nov,
       volume = {705},
       number = {2},
        pages = {1129-1138},
          doi = {10.1088/0004-637X/705/2/1129},
archivePrefix = {arXiv},
       eprint = {0903.4895},
 primaryClass = {astro-ph.CO},
       adsurl = {https://ui.adsabs.harvard.edu/abs/2009ApJ...705.1129L}
}

@ARTICLE{2009A&A...504...33V,
       author = {{Vazza}, F. and {Brunetti}, G. and {Kritsuk}, A. and {Wagner}, R. and {Gheller}, C. and {Norman}, M.},
        title = "{Turbulent motions and shocks waves in galaxy clusters simulated with adaptive mesh refinement}",
      journal = {Astronomy \& Astrophysics},
         year = 2009,
        month = sep,
       volume = {504},
       number = {1},
        pages = {33-43},
          doi = {10.1051/0004-6361/200912535},
archivePrefix = {arXiv},
       eprint = {0905.3169},
 primaryClass = {astro-ph.CO},
       adsurl = {https://ui.adsabs.harvard.edu/abs/2009A&A...504...33V}
}

@ARTICLE{2012ApJ...758...74B,
       author = {{Battaglia}, N. and {Bond}, J.~R. and {Pfrommer}, C. and {Sievers}, J.~L.},
        title = "{On the Cluster Physics of Sunyaev-Zel'dovich and X-Ray Surveys. I. The Influence of Feedback, Non-thermal Pressure, and Cluster Shapes on Y-M Scaling Relations}",
      journal = {The Astrophysical Journal},
         year = 2012,
        month = oct,
       volume = {758},
       number = {2},
          eid = {74},
        pages = {74},
          doi = {10.1088/0004-637X/758/2/74},
archivePrefix = {arXiv},
       eprint = {1109.3709},
 primaryClass = {astro-ph.CO},
       adsurl = {https://ui.adsabs.harvard.edu/abs/2012ApJ...758...74B}
}

@ARTICLE{2014ApJ...792...25N,
       author = {{Nelson}, Kaylea and {Lau}, Erwin T. and {Nagai}, Daisuke},
        title = "{Hydrodynamic Simulation of Non-thermal Pressure Profiles of Galaxy Clusters}",
      journal = {The Astrophysical Journal},
         year = 2014,
        month = sep,
       volume = {792},
       number = {1},
          eid = {25},
        pages = {25},
          doi = {10.1088/0004-637X/792/1/25},
archivePrefix = {arXiv},
       eprint = {1404.4636},
 primaryClass = {astro-ph.CO},
       adsurl = {https://ui.adsabs.harvard.edu/abs/2014ApJ...792...25N}
}

@ARTICLE{2015MNRAS.448.1020S,
       author = {{Shi}, Xun and {Komatsu}, Eiichiro and {Nelson}, Kaylea and {Nagai}, Daisuke},
        title = "{Analytical model for non-thermal pressure in galaxy clusters - II. Comparison with cosmological hydrodynamics simulation}",
      journal = {Monthly Notices of the Royal Astronomical Society},
         year = 2015,
        month = mar,
       volume = {448},
       number = {1},
        pages = {1020-1029},
          doi = {10.1093/mnras/stv036},
archivePrefix = {arXiv},
       eprint = {1408.3832},
 primaryClass = {astro-ph.CO},
       adsurl = {https://ui.adsabs.harvard.edu/abs/2015MNRAS.448.1020S}
}

@ARTICLE{2018MNRAS.480.2898C,
       author = {{Cui}, Weiguang and {Knebe}, Alexander and {Yepes}, Gustavo and {Pearce}, Frazer and {Power}, Chris and {Dave}, Romeel and {Arth}, Alexander and {Borgani}, Stefano and {Dolag}, Klaus and {Elahi}, Pascal and {Mostoghiu}, Robert and {Murante}, Giuseppe and {Rasia}, Elena and {Stoppacher}, Doris and {Vega-Ferrero}, Jesus and {Wang}, Yang and {Yang}, Xiaohu and {Benson}, Andrew and {Cora}, Sof{\'\i}a A. and {Croton}, Darren J. and {Sinha}, Manodeep and {Stevens}, Adam R.~H. and {Vega-Mart{\'\i}nez}, Cristian A. and {Arthur}, Jake and {Baldi}, Anna S. and {Ca{\~n}as}, Rodrigo and {Cialone}, Giammarco and {Cunnama}, Daniel and {De Petris}, Marco and {Durando}, Giacomo and {Ettori}, Stefano and {Gottl{\"o}ber}, Stefan and {Nuza}, Sebasti{\'a}n E. and {Old}, Lyndsay J. and {Pilipenko}, Sergey and {Sorce}, Jenny G. and {Welker}, Charlotte},
        title = "{The Three Hundred project: a large catalogue of theoretically modelled galaxy clusters for cosmological and astrophysical applications}",
      journal = {Monthly Notices of the Royal Astronomical Society},
         year = 2018,
        month = nov,
       volume = {480},
       number = {3},
        pages = {2898-2915},
          doi = {10.1093/mnras/sty2111},
archivePrefix = {arXiv},
       eprint = {1809.04622},
 primaryClass = {astro-ph.GA},
       adsurl = {https://ui.adsabs.harvard.edu/abs/2018MNRAS.480.2898C}
}

@ARTICLE{2015A&A...575A..30S,
       author = {{Schellenberger}, G. and {Reiprich}, T.~H. and {Lovisari}, L. and {Nevalainen}, J. and {David}, L.},
        title = "{XMM-Newton and Chandra cross-calibration using HIFLUGCS galaxy clusters . Systematic temperature differences and cosmological impact}",
      journal = {Astronomy \& Astrophysics},
         year = 2015,
        month = mar,
       volume = {575},
          eid = {A30},
        pages = {A30},
          doi = {10.1051/0004-6361/201424085},
archivePrefix = {arXiv},
       eprint = {1404.7130},
 primaryClass = {astro-ph.CO},
       adsurl = {https://ui.adsabs.harvard.edu/abs/2015A&A...575A..30S}
}

@ARTICLE{2021MNRAS.502.5115G,
       author = {{Gianfagna}, Giulia and {De Petris}, Marco and {Yepes}, Gustavo and {De Luca}, Federico and {Sembolini}, Federico and {Cui}, Weiguang and {Biffi}, Veronica and {K{\'e}ruzor{\'e}}, Florian and {Mac{\'\i}as-P{\'e}rez}, Juan and {Mayet}, Fr{\'e}d{\'e}ric and {Perotto}, Laurence and {Rasia}, Elena and {Ruppin}, Florian},
        title = "{Exploring the hydrostatic mass bias in MUSIC clusters: application to the NIKA2 mock sample}",
      journal = {Monthly Notices of the Royal Astronomical Society},
         year = 2021,
        month = apr,
       volume = {502},
       number = {4},
        pages = {5115-5133},
          doi = {10.1093/mnras/stab308},
archivePrefix = {arXiv},
       eprint = {2010.03634},
 primaryClass = {astro-ph.CO},
       adsurl = {https://ui.adsabs.harvard.edu/abs/2021MNRAS.502.5115G}
}

@ARTICLE{2012NJPh...14e5018R,
       author = {{Rasia}, E. and {Meneghetti}, M. and {Martino}, R. and {Borgani}, S. and {Bonafede}, A. and {Dolag}, K. and {Ettori}, S. and {Fabjan}, D. and {Giocoli}, C. and {Mazzotta}, P. and {Merten}, J. and {Radovich}, M. and {Tornatore}, L.},
        title = "{Lensing and x-ray mass estimates of clusters (simulations)}",
      journal = {New Journal of Physics},
         year = 2012,
        month = may,
       volume = {14},
       number = {5},
          eid = {055018},
        pages = {055018},
          doi = {10.1088/1367-2630/14/5/055018},
archivePrefix = {arXiv},
       eprint = {1201.1569},
 primaryClass = {astro-ph.CO},
       adsurl = {https://ui.adsabs.harvard.edu/abs/2012NJPh...14e5018R}
}

@ARTICLE{1972CoASP...4..173S,
       author = {{Sunyaev}, R.~A. and {Zeldovich}, Ya. B.},
        title = "{The Observations of Relic Radiation as a Test of the Nature of X-Ray Radiation from the Clusters of Galaxies}",
      journal = {Comments on Astrophysics and Space Physics},
         year = 1972,
        month = nov,
       volume = {4},
        pages = {173},
       adsurl = {https://ui.adsabs.harvard.edu/abs/1972CoASP...4..173S}
}

@ARTICLE{2016A&A...594A..27P,
       author = {{Planck Collaboration} and {Ade}, P.~A.~R. and {Aghanim}, N. and {Arnaud}, M. and {Ashdown}, M. and {Aumont}, J. and {Baccigalupi}, C. and {Banday}, A.~J. and {Barreiro}, R.~B. and {Barrena}, R. and {Bartlett}, J.~G. and {Bartolo}, N. and {Battaner}, E. and {Battye}, R. and {Benabed}, K. and {Beno{\^\i}t}, A. and {Benoit-L{\'e}vy}, A. and {Bernard}, J. -P. and {Bersanelli}, M. and {Bielewicz}, P. and {Bikmaev}, I. and {B{\"o}hringer}, H. and {Bonaldi}, A. and {Bonavera}, L. and {Bond}, J.~R. and {Borrill}, J. and {Bouchet}, F.~R. and {Bucher}, M. and {Burenin}, R. and {Burigana}, C. and {Butler}, R.~C. and {Calabrese}, E. and {Cardoso}, J. -F. and {Carvalho}, P. and {Catalano}, A. and {Challinor}, A. and {Chamballu}, A. and {Chary}, R. -R. and {Chiang}, H.~C. and {Chon}, G. and {Christensen}, P.~R. and {Clements}, D.~L. and {Colombi}, S. and {Colombo}, L.~P.~L. and {Combet}, C. and {Comis}, B. and {Couchot}, F. and {Coulais}, A. and {Crill}, B.~P. and {Curto}, A. and {Cuttaia}, F. and {Dahle}, H. and {Danese}, L. and {Davies}, R.~D. and {Davis}, R.~J. and {de Bernardis}, P. and {de Rosa}, A. and {de Zotti}, G. and {Delabrouille}, J. and {D{\'e}sert}, F. -X. and {Dickinson}, C. and {Diego}, J.~M. and {Dolag}, K. and {Dole}, H. and {Donzelli}, S. and {Dor{\'e}}, O. and {Douspis}, M. and {Ducout}, A. and {Dupac}, X. and {Efstathiou}, G. and {Eisenhardt}, P.~R.~M. and {Elsner}, F. and {En{\ss}lin}, T.~A. and {Eriksen}, H.~K. and {Falgarone}, E. and {Fergusson}, J. and {Feroz}, F. and {Ferragamo}, A. and {Finelli}, F. and {Forni}, O. and {Frailis}, M. and {Fraisse}, A.~A. and {Franceschi}, E. and {Frejsel}, A. and {Galeotta}, S. and {Galli}, S. and {Ganga}, K. and {G{\'e}nova-Santos}, R.~T. and {Giard}, M. and {Giraud-H{\'e}raud}, Y. and {Gjerl{\o}w}, E. and {Gonz{\'a}lez-Nuevo}, J. and {G{\'o}rski}, K.~M. and {Grainge}, K.~J.~B. and {Gratton}, S. and {Gregorio}, A. and {Gruppuso}, A. and {Gudmundsson}, J.~E. and {Hansen}, F.~K. and {Hanson}, D. and {Harrison}, D.~L. and {Hempel}, A. and {Henrot-Versill{\'e}}, S. and {Hern{\'a}ndez-Monteagudo}, C. and {Herranz}, D. and {Hildebrandt}, S.~R. and {Hivon}, E. and {Hobson}, M. and {Holmes}, W.~A. and {Hornstrup}, A. and {Hovest}, W. and {Huffenberger}, K.~M. and {Hurier}, G. and {Jaffe}, A.~H. and {Jaffe}, T.~R. and {Jin}, T. and {Jones}, W.~C. and {Juvela}, M. and {Keih{\"a}nen}, E. and {Keskitalo}, R. and {Khamitov}, I. and {Kisner}, T.~S. and {Kneissl}, R. and {Knoche}, J. and {Kunz}, M. and {Kurki-Suonio}, H. and {Lagache}, G. and {Lamarre}, J. -M. and {Lasenby}, A. and {Lattanzi}, M. and {Lawrence}, C.~R. and {Leonardi}, R. and {Lesgourgues}, J. and {Levrier}, F. and {Liguori}, M. and {Lilje}, P.~B. and {Linden-V{\o}rnle}, M. and {L{\'o}pez-Caniego}, M. and {Lubin}, P.~M. and {Mac{\'\i}as-P{\'e}rez}, J.~F. and {Maggio}, G. and {Maino}, D. and {Mak}, D.~S.~Y. and {Mandolesi}, N. and {Mangilli}, A. and {Martin}, P.~G. and {Mart{\'\i}nez-Gonz{\'a}lez}, E. and {Masi}, S. and {Matarrese}, S. and {Mazzotta}, P. and {McGehee}, P. and {Mei}, S. and {Melchiorri}, A. and {Melin}, J. -B. and {Mendes}, L. and {Mennella}, A. and {Migliaccio}, M. and {Mitra}, S. and {Miville-Desch{\^e}nes}, M. -A. and {Moneti}, A. and {Montier}, L. and {Morgante}, G. and {Mortlock}, D. and {Moss}, A. and {Munshi}, D. and {Murphy}, J.~A. and {Naselsky}, P. and {Nastasi}, A. and {Nati}, F. and {Natoli}, P. and {Netterfield}, C.~B. and {N{\o}rgaard-Nielsen}, H.~U. and {Noviello}, F. and {Novikov}, D. and {Novikov}, I. and {Olamaie}, M. and {Oxborrow}, C.~A. and {Paci}, F. and {Pagano}, L. and {Pajot}, F. and {Paoletti}, D. and {Pasian}, F. and {Patanchon}, G. and {Pearson}, T.~J. and {Perdereau}, O. and {Perotto}, L. and {Perrott}, Y.~C. and {Perrotta}, F. and {Pettorino}, V. and {Piacentini}, F. and {Piat}, M. and {Pierpaoli}, E. and {Pietrobon}, D. and {Plaszczynski}, S. and {Pointecouteau}, E. and {Polenta}, G. and {Pratt}, G.~W. and {Pr{\'e}zeau}, G. and {Prunet}, S. and {Puget}, J. -L. and {Rachen}, J.~P. and {Reach}, W.~T. and {Rebolo}, R. and {Reinecke}, M. and {Remazeilles}, M. and {Renault}, C. and {Renzi}, A. and {Ristorcelli}, I. and {Rocha}, G. and {Rosset}, C. and {Rossetti}, M. and {Roudier}, G. and {Rozo}, E. and {Rubi{\~n}o-Mart{\'\i}n}, J.~A. and {Rumsey}, C. and {Rusholme}, B. and {Rykoff}, E.~S. and {Sandri}, M. and {Santos}, D. and {Saunders}, R.~D.~E. and {Savelainen}, M. and {Savini}, G. and {Schammel}, M.~P. and {Scott}, D. and {Seiffert}, M.~D. and {Shellard}, E.~P.~S. and {Shimwell}, T.~W. and {Spencer}, L.~D. and {Stanford}, S.~A. and {Stern}, D. and {Stolyarov}, V. and {Stompor}, R. and {Streblyanska}, A. and {Sudiwala}, R. and {Sunyaev}, R. and {Sutton}, D. and {Suur-Uski}, A. -S. and {Sygnet}, J. -F. and {Tauber}, J.~A. and {Terenzi}, L. and {Toffolatti}, L. and {Tomasi}, M. and {Tramonte}, D. and {Tristram}, M. and {Tucci}, M. and {Tuovinen}, J. and {Umana}, G. and {Valenziano}, L. and {Valiviita}, J. and {Van Tent}, B. and {Vielva}, P. and {Villa}, F. and {Wade}, L.~A. and {Wandelt}, B.~D. and {Wehus}, I.~K. and {White}, S.~D.~M. and {Wright}, E.~L. and {Yvon}, D. and {Zacchei}, A. and {Zonca}, A.},
        title = "{Planck 2015 results. XXVII. The second Planck catalogue of Sunyaev-Zeldovich sources}",
      journal = {Astronomy \& Astrophysics},
         year = 2016,
        month = sep,
       volume = {594},
          eid = {A27},
        pages = {A27},
          doi = {10.1051/0004-6361/201525823},
archivePrefix = {arXiv},
       eprint = {1502.01598},
 primaryClass = {astro-ph.CO},
       adsurl = {https://ui.adsabs.harvard.edu/abs/2016A&A...594A..27P}
}

@ARTICLE{2023A&A...674A..48W,
       author = {{Wicker}, R. and {Douspis}, M. and {Salvati}, L. and {Aghanim}, N.},
        title = "{Constraining the mass and redshift evolution of the hydrostatic mass bias using the gas mass fraction in galaxy clusters}",
      journal = {Astronomy \& Astrophysics},
         year = 2023,
        month = jun,
       volume = {674},
          eid = {A48},
        pages = {A48},
          doi = {10.1051/0004-6361/202243922},
archivePrefix = {arXiv},
       eprint = {2204.12823},
 primaryClass = {astro-ph.CO},
       adsurl = {https://ui.adsabs.harvard.edu/abs/2023A&A...674A..48W}
}

@BOOK{1988xrec.book.....S,
       author = {{Sarazin}, Craig L.},
        title = "{X-ray emission from clusters of galaxies}",
         year = 1988,
       adsurl = {https://ui.adsabs.harvard.edu/abs/1988xrec.book.....S}
}

@ARTICLE{2016A&A...594A..13P,
       author = {{Planck Collaboration} and {Ade}, P.~A.~R. and {Aghanim}, N. and {Arnaud}, M. and {Ashdown}, M. and {Aumont}, J. and {Baccigalupi}, C. and {Banday}, A.~J. and {Barreiro}, R.~B. and {Bartlett}, J.~G. and {Bartolo}, N. and {Battaner}, E. and {Battye}, R. and {Benabed}, K. and {Beno{\^\i}t}, A. and {Benoit-L{\'e}vy}, A. and {Bernard}, J. -P. and {Bersanelli}, M. and {Bielewicz}, P. and {Bock}, J.~J. and {Bonaldi}, A. and {Bonavera}, L. and {Bond}, J.~R. and {Borrill}, J. and {Bouchet}, F.~R. and {Boulanger}, F. and {Bucher}, M. and {Burigana}, C. and {Butler}, R.~C. and {Calabrese}, E. and {Cardoso}, J. -F. and {Catalano}, A. and {Challinor}, A. and {Chamballu}, A. and {Chary}, R. -R. and {Chiang}, H.~C. and {Chluba}, J. and {Christensen}, P.~R. and {Church}, S. and {Clements}, D.~L. and {Colombi}, S. and {Colombo}, L.~P.~L. and {Combet}, C. and {Coulais}, A. and {Crill}, B.~P. and {Curto}, A. and {Cuttaia}, F. and {Danese}, L. and {Davies}, R.~D. and {Davis}, R.~J. and {de Bernardis}, P. and {de Rosa}, A. and {de Zotti}, G. and {Delabrouille}, J. and {D{\'e}sert}, F. -X. and {Di Valentino}, E. and {Dickinson}, C. and {Diego}, J.~M. and {Dolag}, K. and {Dole}, H. and {Donzelli}, S. and {Dor{\'e}}, O. and {Douspis}, M. and {Ducout}, A. and {Dunkley}, J. and {Dupac}, X. and {Efstathiou}, G. and {Elsner}, F. and {En{\ss}lin}, T.~A. and {Eriksen}, H.~K. and {Farhang}, M. and {Fergusson}, J. and {Finelli}, F. and {Forni}, O. and {Frailis}, M. and {Fraisse}, A.~A. and {Franceschi}, E. and {Frejsel}, A. and {Galeotta}, S. and {Galli}, S. and {Ganga}, K. and {Gauthier}, C. and {Gerbino}, M. and {Ghosh}, T. and {Giard}, M. and {Giraud-H{\'e}raud}, Y. and {Giusarma}, E. and {Gjerl{\o}w}, E. and {Gonz{\'a}lez-Nuevo}, J. and {G{\'o}rski}, K.~M. and {Gratton}, S. and {Gregorio}, A. and {Gruppuso}, A. and {Gudmundsson}, J.~E. and {Hamann}, J. and {Hansen}, F.~K. and {Hanson}, D. and {Harrison}, D.~L. and {Helou}, G. and {Henrot-Versill{\'e}}, S. and {Hern{\'a}ndez-Monteagudo}, C. and {Herranz}, D. and {Hildebrandt}, S.~R. and {Hivon}, E. and {Hobson}, M. and {Holmes}, W.~A. and {Hornstrup}, A. and {Hovest}, W. and {Huang}, Z. and {Huffenberger}, K.~M. and {Hurier}, G. and {Jaffe}, A.~H. and {Jaffe}, T.~R. and {Jones}, W.~C. and {Juvela}, M. and {Keih{\"a}nen}, E. and {Keskitalo}, R. and {Kisner}, T.~S. and {Kneissl}, R. and {Knoche}, J. and {Knox}, L. and {Kunz}, M. and {Kurki-Suonio}, H. and {Lagache}, G. and {L{\"a}hteenm{\"a}ki}, A. and {Lamarre}, J. -M. and {Lasenby}, A. and {Lattanzi}, M. and {Lawrence}, C.~R. and {Leahy}, J.~P. and {Leonardi}, R. and {Lesgourgues}, J. and {Levrier}, F. and {Lewis}, A. and {Liguori}, M. and {Lilje}, P.~B. and {Linden-V{\o}rnle}, M. and {L{\'o}pez-Caniego}, M. and {Lubin}, P.~M. and {Mac{\'\i}as-P{\'e}rez}, J.~F. and {Maggio}, G. and {Maino}, D. and {Mandolesi}, N. and {Mangilli}, A. and {Marchini}, A. and {Maris}, M. and {Martin}, P.~G. and {Martinelli}, M. and {Mart{\'\i}nez-Gonz{\'a}lez}, E. and {Masi}, S. and {Matarrese}, S. and {McGehee}, P. and {Meinhold}, P.~R. and {Melchiorri}, A. and {Melin}, J. -B. and {Mendes}, L. and {Mennella}, A. and {Migliaccio}, M. and {Millea}, M. and {Mitra}, S. and {Miville-Desch{\^e}nes}, M. -A. and {Moneti}, A. and {Montier}, L. and {Morgante}, G. and {Mortlock}, D. and {Moss}, A. and {Munshi}, D. and {Murphy}, J.~A. and {Naselsky}, P. and {Nati}, F. and {Natoli}, P. and {Netterfield}, C.~B. and {N{\o}rgaard-Nielsen}, H.~U. and {Noviello}, F. and {Novikov}, D. and {Novikov}, I. and {Oxborrow}, C.~A. and {Paci}, F. and {Pagano}, L. and {Pajot}, F. and {Paladini}, R. and {Paoletti}, D. and {Partridge}, B. and {Pasian}, F. and {Patanchon}, G. and {Pearson}, T.~J. and {Perdereau}, O. and {Perotto}, L. and {Perrotta}, F. and {Pettorino}, V. and {Piacentini}, F. and {Piat}, M. and {Pierpaoli}, E. and {Pietrobon}, D. and {Plaszczynski}, S. and {Pointecouteau}, E. and {Polenta}, G. and {Popa}, L. and {Pratt}, G.~W. and {Pr{\'e}zeau}, G. and {Prunet}, S. and {Puget}, J. -L. and {Rachen}, J.~P. and {Reach}, W.~T. and {Rebolo}, R. and {Reinecke}, M. and {Remazeilles}, M. and {Renault}, C. and {Renzi}, A. and {Ristorcelli}, I. and {Rocha}, G. and {Rosset}, C. and {Rossetti}, M. and {Roudier}, G. and {Rouill{\'e} d'Orfeuil}, B. and {Rowan-Robinson}, M. and {Rubi{\~n}o-Mart{\'\i}n}, J.~A. and {Rusholme}, B. and {Said}, N. and {Salvatelli}, V. and {Salvati}, L. and {Sandri}, M. and {Santos}, D. and {Savelainen}, M. and {Savini}, G. and {Scott}, D. and {Seiffert}, M.~D. and {Serra}, P. and {Shellard}, E.~P.~S. and {Spencer}, L.~D. and {Spinelli}, M. and {Stolyarov}, V. and {Stompor}, R. and {Sudiwala}, R. and {Sunyaev}, R. and {Sutton}, D. and {Suur-Uski}, A. -S. and {Sygnet}, J. -F. and {Tauber}, J.~A. and {Terenzi}, L. and {Toffolatti}, L. and {Tomasi}, M. and {Tristram}, M. and {Trombetti}, T. and {Tucci}, M. and {Tuovinen}, J. and {T{\"u}rler}, M. and {Umana}, G. and {Valenziano}, L. and {Valiviita}, J. and {Van Tent}, F. and {Vielva}, P. and {Villa}, F. and {Wade}, L.~A. and {Wandelt}, B.~D. and {Wehus}, I.~K. and {White}, M. and {White}, S.~D.~M. and {Wilkinson}, A. and {Yvon}, D. and {Zacchei}, A. and {Zonca}, A.},
        title = "{Planck 2015 results. XIII. Cosmological parameters}",
      journal = {Astronomy \& Astrophysics},
         year = 2016,
        month = sep,
       volume = {594},
          eid = {A13},
        pages = {A13},
          doi = {10.1051/0004-6361/201525830},
archivePrefix = {arXiv},
       eprint = {1502.01589},
 primaryClass = {astro-ph.CO},
       adsurl = {https://ui.adsabs.harvard.edu/abs/2016A&A...594A..13P}
}

@ARTICLE{2007PhR...443....1M,
       author = {{Markevitch}, Maxim and {Vikhlinin}, Alexey},
        title = "{Shocks and cold fronts in galaxy clusters}",
      journal = {Physics Report},
         year = 2007,
        month = may,
       volume = {443},
       number = {1},
        pages = {1-53},
          doi = {10.1016/j.physrep.2007.01.001},
archivePrefix = {arXiv},
       eprint = {astro-ph/0701821},
 primaryClass = {astro-ph},
       adsurl = {https://ui.adsabs.harvard.edu/abs/2007PhR...443....1M}
}

@ARTICLE{2024A&A...688A.107M,
       author = {{Migkas}, K. and {Kox}, D. and {Schellenberger}, G. and {Veronica}, A. and {Pacaud}, F. and {Reiprich}, T.~H. and {Bahar}, Y.~E. and {Balzer}, F. and {Bulbul}, E. and {Comparat}, J. and {Dennerl}, K. and {Freyberg}, M. and {Garrel}, C. and {Ghirardini}, V. and {Grandis}, S. and {Kluge}, M. and {Liu}, A. and {Ramos-Ceja}, M.~E. and {Sanders}, J. and {Zhang}, X.},
        title = "{The SRG/eROSITA All-Sky Survey. SRG/eROSITA cross-calibration with Chandra and XMM-Newton using galaxy cluster gas temperatures}",
      journal = {A\&A},
         year = 2024,
        month = aug,
       volume = {688},
          eid = {A107},
        pages = {A107},
          doi = {10.1051/0004-6361/202349006},
archivePrefix = {arXiv},
       eprint = {2401.17297},
 primaryClass = {astro-ph.CO},
       adsurl = {https://ui.adsabs.harvard.edu/abs/2024A&A...688A.107M}
}

@ARTICLE{2017A&A...606A..64A,
       author = {{Adam}, R. and {Arnaud}, M. and {Bartalucci}, I. and {Ade}, P. and {Andr{\'e}}, P. and {Beelen}, A. and {Beno{\^\i}t}, A. and {Bideaud}, A. and {Billot}, N. and {Bourdin}, H. and {Bourrion}, O. and {Calvo}, M. and {Catalano}, A. and {Coiffard}, G. and {Comis}, B. and {D'Addabbo}, A. and {D{\'e}sert}, F. -X. and {Doyle}, S. and {Ferrari}, C. and {Goupy}, J. and {Kramer}, C. and {Lagache}, G. and {Leclercq}, S. and {Mac{\'\i}as-P{\'e}rez}, J. -F. and {Maurogordato}, S. and {Mauskopf}, P. and {Mayet}, F. and {Monfardini}, A. and {Pajot}, F. and {Pascale}, E. and {Perotto}, L. and {Pisano}, G. and {Pointecouteau}, E. and {Ponthieu}, N. and {Pratt}, G.~W. and {Rev{\'e}ret}, V. and {Ritacco}, A. and {Rodriguez}, L. and {Romero}, C. and {Ruppin}, F. and {Schuster}, K. and {Sievers}, A. and {Triqueneaux}, S. and {Tucker}, C. and {Zylka}, R.},
        title = "{Mapping the hot gas temperature in galaxy clusters using X-ray and Sunyaev-Zel'dovich imaging}",
      journal = {A\&A},
         year = 2017,
        month = oct,
       volume = {606},
          eid = {A64},
        pages = {A64},
          doi = {10.1051/0004-6361/201629810},
archivePrefix = {arXiv},
       eprint = {1706.10230},
 primaryClass = {astro-ph.CO},
       adsurl = {https://ui.adsabs.harvard.edu/abs/2017A&A...606A..64A}
}

@INPROCEEDINGS{2022EPJWC.25700036P,
       author = {{Paliwal}, A. and {Artis}, E. and {Cui}, W. and {De Petris}, M. and {D{\'e}sert}, F. -X. and {Ferragamo}, A. and {Gianfagna}, G. and {K{\'e}ruzor{\'e}}, F. and {Mac{\'\i}as-P{\'e}rez}, J. -F. and {Mayet}, F. and {Mu{\~n}oz-Echeverr{\'\i}a}, M. and {Perotto}, L. and {Rasia}, E. and {Ruppin}, F. and {Yepes}, G.},
        title = "{The Three Hundred-NIKA2 Sunyaev-Zeldovich Large Program twin samples: Synthetic clusters to support real observations}",
    booktitle = {mm Universe @ NIKA2 - Observing the mm Universe with the NIKA2 Camera},
         year = 2022,
       series = {European Physical Journal Web of Conferences},
       volume = {257},
        month = jul,
          eid = {00036},
        pages = {00036},
          doi = {10.1051/epjconf/202225700036},
archivePrefix = {arXiv},
       eprint = {2111.01920},
 primaryClass = {astro-ph.CO},
       adsurl = {https://ui.adsabs.harvard.edu/abs/2022EPJWC.25700036P}
}

@INPROCEEDINGS{2020EPJWC.22800017M,
       author = {{Mayet}, F. and {Adam}, R. and {Ade}, P. and {Andr{\'e}}, P. and {Andrianasolo}, A. and {Arnaud}, M. and {Aussel}, H. and {Bartalucci}, I. and {Beelen}, A. and {Beno{\^\i}t}, A. and {Bideaud}, A. and {Bourrion}, O. and {Calvo}, M. and {Catalano}, A. and {Comis}, B. and {De Petris}, M. and {D{\'e}sert}, F. -X. and {Doyle}, S. and {Driessen}, E.~F.~C. and {Gomez}, A. and {Goupy}, J. and {K{\'e}ruzor{\'e}}, F. and {Kramer}, C. and {Ladjelate}, B. and {Lagache}, G. and {Leclercq}, S. and {Lestrade}, J. -F. and {Mac{\'\i}as-P{\'e}rez}, J.~F. and {Mauskopf}, P. and {Monfardini}, A. and {Perotto}, L. and {Pisano}, G. and {Pointecouteau}, E. and {Ponthieu}, N. and {Pratt}, G.~W. and {Rev{\'e}ret}, V. and {Ritacco}, A. and {Romero}, C. and {Roussel}, H. and {Ruppin}, F. and {Schuster}, K. and {Shu}, S. and {Sievers}, A. and {Tucker}, C. and {Zylka}, R.},
        title = "{Cluster cosmology with the NIKA2 SZ Large Program}",
    booktitle = {mm Universe @ NIKA2 - Observing the mm Universe with the NIKA2 Camera},
         year = 2020,
       series = {European Physical Journal Web of Conferences},
       volume = {228},
        month = jun,
          eid = {00017},
        pages = {00017},
          doi = {10.1051/epjconf/202022800017},
archivePrefix = {arXiv},
       eprint = {1911.03145},
 primaryClass = {astro-ph.CO},
       adsurl = {https://ui.adsabs.harvard.edu/abs/2020EPJWC.22800017M}
}

@ARTICLE{2012MNRAS.426..510C,
       author = {{Chluba}, Jens and {Nagai}, Daisuke and {Sazonov}, Sergey and {Nelson}, Kaylea},
        title = "{A fast and accurate method for computing the Sunyaev-Zel'dovich signal of hot galaxy clusters}",
      journal = {MNRAS},
         year = 2012,
        month = oct,
       volume = {426},
       number = {1},
        pages = {510-530},
          doi = {10.1111/j.1365-2966.2012.21741.x},
archivePrefix = {arXiv},
       eprint = {1205.5778},
 primaryClass = {astro-ph.CO},
       adsurl = {https://ui.adsabs.harvard.edu/abs/2012MNRAS.426..510C}
}

@ARTICLE{2013MNRAS.430.3054C,
       author = {{Chluba}, Jens and {Switzer}, Eric and {Nelson}, Kaylea and {Nagai}, Daisuke},
        title = "{Sunyaev-Zeldovich signal processing and temperature-velocity moment method for individual clusters}",
      journal = {MNRAS},
         year = 2013,
        month = apr,
       volume = {430},
       number = {4},
        pages = {3054-3069},
          doi = {10.1093/mnras/stt110},
archivePrefix = {arXiv},
       eprint = {1211.3206},
 primaryClass = {astro-ph.CO},
       adsurl = {https://ui.adsabs.harvard.edu/abs/2013MNRAS.430.3054C}
}

@ARTICLE{2007PASA...24..159P,
       author = {{Price}, Daniel J.},
        title = "{splash: An Interactive Visualisation Tool for Smoothed Particle Hydrodynamics Simulations}",
      journal = {\pasa},
         year = 2007,
        month = oct,
       volume = {24},
       number = {3},
        pages = {159-173},
          doi = {10.1071/AS07022},
archivePrefix = {arXiv},
       eprint = {0709.0832},
 primaryClass = {astro-ph},
       adsurl = {https://ui.adsabs.harvard.edu/abs/2007PASA...24..159P}
}

@ARTICLE{2014MNRAS.439..588B,
       author = {{Biffi}, V. and {Sembolini}, F. and {De Petris}, M. and {Valdarnini}, R. and {Yepes}, G. and {Gottl{\"o}ber}, S.},
        title = "{The MUSIC of galaxy clusters - II. X-ray global properties and scaling relations}",
      journal = {\mnras},
         year = 2014,
        month = mar,
       volume = {439},
       number = {1},
        pages = {588-603},
          doi = {10.1093/mnras/stu018},
archivePrefix = {arXiv},
       eprint = {1401.2992},
 primaryClass = {astro-ph.CO},
       adsurl = {https://ui.adsabs.harvard.edu/abs/2014MNRAS.439..588B}
}

@ARTICLE{2004MNRAS.354...10M,
       author = {{Mazzotta}, P. and {Rasia}, E. and {Moscardini}, L. and {Tormen}, G.},
        title = "{Comparing the temperatures of galaxy clusters from hydrodynamical N-body simulations to Chandra and XMM-Newton observations}",
      journal = {\mnras},
         year = 2004,
        month = oct,
       volume = {354},
       number = {1},
        pages = {10-24},
          doi = {10.1111/j.1365-2966.2004.08167.x},
archivePrefix = {arXiv},
       eprint = {astro-ph/0404425},
 primaryClass = {astro-ph},
       adsurl = {https://ui.adsabs.harvard.edu/abs/2004MNRAS.354...10M}
}

@ARTICLE{2023A&A...675A.150Z,
       author = {{ZuHone}, J. and {Bahar}, Y.~E. and {Biffi}, V. and {Dolag}, K. and {Sanders}, J. and {Bulbul}, E. and {Liu}, T. and {Dauser}, T. and {K{\"o}nig}, O. and {Zhang}, X. and {Ghirardini}, V.},
        title = "{Effects of multiphase gas and projection on X-ray observables in simulated galaxy clusters as seen by eROSITA}",
      journal = {Astronomy \& Astrophysics},
         year = 2023,
        month = jul,
       volume = {675},
          eid = {A150},
        pages = {A150},
          doi = {10.1051/0004-6361/202245749},
archivePrefix = {arXiv},
       eprint = {2212.11028},
 primaryClass = {astro-ph.CO},
       adsurl = {https://ui.adsabs.harvard.edu/abs/2023A&A...675A.150Z}
}

@ARTICLE{2023MNRAS.522..721P,
       author = {{Pellissier}, Alisson and {Hahn}, Oliver and {Ferrari}, Chiara},
        title = "{RHAPSODY-C simulations - anisotropic thermal conduction, black hole physics, and the robustness of massive galaxy cluster scaling relations}",
      journal = {\mnras},
         year = 2023,
        month = jun,
       volume = {522},
       number = {1},
        pages = {721-749},
          doi = {10.1093/mnras/stad888},
archivePrefix = {arXiv},
       eprint = {2301.02684},
 primaryClass = {astro-ph.CO},
       adsurl = {https://ui.adsabs.harvard.edu/abs/2023MNRAS.522..721P}
}

@ARTICLE{2019MNRAS.489.2439H,
       author = {{Henden}, Nicholas A. and {Puchwein}, Ewald and {Sijacki}, Debora},
        title = "{The redshift evolution of X-ray and Sunyaev-Zel'dovich scaling relations in the FABLE simulations}",
      journal = {\mnras},
         year = 2019,
        month = oct,
       volume = {489},
       number = {2},
        pages = {2439-2470},
          doi = {10.1093/mnras/stz2301},
archivePrefix = {arXiv},
       eprint = {1905.00013},
 primaryClass = {astro-ph.CO},
       adsurl = {https://ui.adsabs.harvard.edu/abs/2019MNRAS.489.2439H}
}

@ARTICLE{2018MNRAS.479.5385H,
       author = {{Henden}, Nicholas A. and {Puchwein}, Ewald and {Shen}, Sijing and {Sijacki}, Debora},
        title = "{The FABLE simulations: a feedback model for galaxies, groups, and clusters}",
      journal = {\mnras},
         year = 2018,
        month = oct,
       volume = {479},
       number = {4},
        pages = {5385-5412},
          doi = {10.1093/mnras/sty1780},
archivePrefix = {arXiv},
       eprint = {1804.05064},
 primaryClass = {astro-ph.GA},
       adsurl = {https://ui.adsabs.harvard.edu/abs/2018MNRAS.479.5385H}
}

@ARTICLE{2017MNRAS.470..166H,
       author = {{Hahn}, Oliver and {Martizzi}, Davide and {Wu}, Hao-Yi and {Evrard}, August E. and {Teyssier}, Romain and {Wechsler}, Risa H.},
        title = "{rhapsody-g simulations - I. The cool cores, hot gas and stellar content of massive galaxy clusters}",
      journal = {\mnras},
         year = 2017,
        month = sep,
       volume = {470},
       number = {1},
        pages = {166-186},
          doi = {10.1093/mnras/stx001},
archivePrefix = {arXiv},
       eprint = {1509.04289},
 primaryClass = {astro-ph.CO},
       adsurl = {https://ui.adsabs.harvard.edu/abs/2017MNRAS.470..166H}
}

@ARTICLE{2021MNRAS.504.5383D,
       author = {{De Luca}, Federico and {De Petris}, Marco and {Yepes}, Gustavo and {Cui}, Weiguang and {Knebe}, Alexander and {Rasia}, Elena},
        title = "{The Three Hundred project: dynamical state of galaxy clusters and morphology from multiwavelength synthetic maps}",
      journal = {\mnras},
         year = 2021,
        month = jul,
       volume = {504},
       number = {4},
        pages = {5383-5400},
          doi = {10.1093/mnras/stab1073},
archivePrefix = {arXiv},
       eprint = {2011.09002},
 primaryClass = {astro-ph.CO},
       adsurl = {https://ui.adsabs.harvard.edu/abs/2021MNRAS.504.5383D}
}

@ARTICLE{2006ApJ...640..691V,
       author = {{Vikhlinin}, A. and {Kravtsov}, A. and {Forman}, W. and {Jones}, C. and {Markevitch}, M. and {Murray}, S.~S. and {Van Speybroeck}, L.},
        title = "{Chandra Sample of Nearby Relaxed Galaxy Clusters: Mass, Gas Fraction, and Mass-Temperature Relation}",
      journal = {\apj},
         year = 2006,
        month = apr,
       volume = {640},
       number = {2},
        pages = {691-709},
          doi = {10.1086/500288},
archivePrefix = {arXiv},
       eprint = {astro-ph/0507092},
 primaryClass = {astro-ph},
       adsurl = {https://ui.adsabs.harvard.edu/abs/2006ApJ...640..691V}
}

@ARTICLE{2005MNRAS.363...29D,
       author = {{Dolag}, K. and {Hansen}, F.~K. and {Roncarelli}, M. and {Moscardini}, L.},
        title = "{The imprints of local superclusters on the Sunyaev-Zel'dovich signals and their detectability with Planck}",
      journal = {\mnras},
         year = 2005,
        month = oct,
       volume = {363},
       number = {1},
        pages = {29-39},
          doi = {10.1111/j.1365-2966.2005.09452.x},
archivePrefix = {arXiv},
       eprint = {astro-ph/0505258},
 primaryClass = {astro-ph},
       adsurl = {https://ui.adsabs.harvard.edu/abs/2005MNRAS.363...29D}
}

@ARTICLE{2020A&A...634A.113A,
       author = {{Ansarifard}, S. and {Rasia}, E. and {Biffi}, V. and {Borgani}, S. and {Cui}, W. and {De Petris}, M. and {Dolag}, K. and {Ettori}, S. and {Movahed}, S.~M.~S. and {Murante}, G. and {Yepes}, G.},
        title = "{The Three Hundred Project: Correcting for the hydrostatic-equilibrium mass bias in X-ray and SZ surveys}",
      journal = {\aap},
         year = 2020,
        month = feb,
       volume = {634},
          eid = {A113},
        pages = {A113},
          doi = {10.1051/0004-6361/201936742},
archivePrefix = {arXiv},
       eprint = {1911.07878},
 primaryClass = {astro-ph.CO},
       adsurl = {https://ui.adsabs.harvard.edu/abs/2020A&A...634A.113A}
}

@ARTICLE{2020MNRAS.492.6074H,
       author = {{Haggar}, Roan and {Gray}, Meghan E. and {Pearce}, Frazer R. and {Knebe}, Alexander and {Cui}, Weiguang and {Mostoghiu}, Robert and {Yepes}, Gustavo},
        title = "{The Three Hundred project: backsplash galaxies in simulations of clusters}",
      journal = {\mnras},
         year = 2020,
        month = mar,
       volume = {492},
       number = {4},
        pages = {6074-6085},
          doi = {10.1093/mnras/staa273},
archivePrefix = {arXiv},
       eprint = {2001.11518},
 primaryClass = {astro-ph.GA},
       adsurl = {https://ui.adsabs.harvard.edu/abs/2020MNRAS.492.6074H}
}

@ARTICLE{1976A&A....49..137C,
       author = {{Cavaliere}, A. and {Fusco-Femiano}, R.},
        title = "{X-rays from hot plasma in clusters of galaxies.}",
      journal = {\aap},
         year = 1976,
        month = may,
       volume = {49},
        pages = {137-144},
       adsurl = {https://ui.adsabs.harvard.edu/abs/1976A&A....49..137C}
}

@ARTICLE{2018A&A...614A...7G,
       author = {{Ghirardini}, V. and {Ettori}, S. and {Eckert}, D. and {Molendi}, S. and {Gastaldello}, F. and {Pointecouteau}, E. and {Hurier}, G. and {Bourdin}, H.},
        title = "{The XMM Cluster Outskirts Project (X-COP): Thermodynamic properties of the intracluster medium out to R$_{200}$ in Abell 2319}",
      journal = {Astronomy \& Astrophysics},
         year = 2018,
        month = jun,
       volume = {614},
          eid = {A7},
        pages = {A7},
          doi = {10.1051/0004-6361/201731748},
archivePrefix = {arXiv},
       eprint = {1708.02954},
 primaryClass = {astro-ph.CO},
       adsurl = {https://ui.adsabs.harvard.edu/abs/2018A&A...614A...7G}
}

@article{Mroczkowski_2009,
doi = {10.1088/0004-637X/694/2/1034},
url = {https://dx.doi.org/10.1088/0004-637X/694/2/1034},
year = {2009},
month = {mar},
publisher = {The American Astronomical Society},
volume = {694},
number = {2},
pages = {1034},
author = {Mroczkowski, Tony and Bonamente, Max and Carlstrom, John E. and Culverhouse, Thomas L. and Greer, Christopher and Hawkins, David and Hennessy, Ryan and Joy, Marshall and Lamb, James W. and Leitch, Erik M. and Loh, Michael and Maughan, Ben and Marrone, Daniel P. and Miller, Amber and Muchovej, Stephen and Nagai, Daisuke and Pryke, Clem and Sharp, Matthew and Woody, David},
title = {APPLICATION OF A SELF-SIMILAR PRESSURE PROFILE TO SUNYAEV–ZEL'DOVICH EFFECT DATA FROM GALAXY CLUSTERS},
journal = {The Astrophysical Journal}
}

@ARTICLE{2017A&A...597A.110R,
       author = {{Ruppin}, F. and {Adam}, R. and {Comis}, B. and {Ade}, P. and {Andr{\'e}}, P. and {Arnaud}, M. and {Beelen}, A. and {Beno{\^\i}t}, A. and {Bideaud}, A. and {Billot}, N. and {Bourrion}, O. and {Calvo}, M. and {Catalano}, A. and {Coiffard}, G. and {D'Addabbo}, A. and {De Petris}, M. and {D{\'e}sert}, F. -X. and {Doyle}, S. and {Goupy}, J. and {Kramer}, C. and {Leclercq}, S. and {Mac{\'\i}as-P{\'e}rez}, J.~F. and {Mauskopf}, P. and {Mayet}, F. and {Monfardini}, A. and {Pajot}, F. and {Pascale}, E. and {Perotto}, L. and {Pisano}, G. and {Pointecouteau}, E. and {Ponthieu}, N. and {Pratt}, G.~W. and {Rev{\'e}ret}, V. and {Ritacco}, A. and {Rodriguez}, L. and {Romero}, C. and {Schuster}, K. and {Sievers}, A. and {Triqueneaux}, S. and {Tucker}, C. and {Zylka}, R.},
        title = "{Non-parametric deprojection of NIKA SZ observations: Pressure distribution in the Planck-discovered cluster PSZ1 G045.85+57.71}",
      journal = {Astronomy \& Astrophysics},
         year = 2017,
        month = jan,
       volume = {597},
          eid = {A110},
        pages = {A110},
          doi = {10.1051/0004-6361/201629405},
archivePrefix = {arXiv},
       eprint = {1607.07679},
 primaryClass = {astro-ph.CO},
       adsurl = {https://ui.adsabs.harvard.edu/abs/2017A&A...597A.110R}
}

@ARTICLE{2015A&A...576A..12A,
       author = {{Adam}, R. and {Comis}, B. and {Mac{\'\i}as-P{\'e}rez}, J. -F. and {Adane}, A. and {Ade}, P. and {Andr{\'e}}, P. and {Beelen}, A. and {Belier}, B. and {Beno{\^\i}t}, A. and {Bideaud}, A. and {Billot}, N. and {Blanquer}, G. and {Bourrion}, O. and {Calvo}, M. and {Catalano}, A. and {Coiffard}, G. and {Cruciani}, A. and {D'Addabbo}, A. and {D{\'e}sert}, F. -X. and {Doyle}, S. and {Goupy}, J. and {Kramer}, C. and {Leclercq}, S. and {Martino}, J. and {Mauskopf}, P. and {Mayet}, F. and {Monfardini}, A. and {Pajot}, F. and {Pascale}, E. and {Perotto}, L. and {Pointecouteau}, E. and {Ponthieu}, N. and {Rev{\'e}ret}, V. and {Ritacco}, A. and {Rodriguez}, L. and {Savini}, G. and {Schuster}, K. and {Sievers}, A. and {Tucker}, C. and {Zylka}, R.},
        title = "{Pressure distribution of the high-redshift cluster of galaxies CL J1226.9+3332 with NIKA}",
      journal = {Astronomy \& Astrophysics},
         year = 2015,
        month = apr,
       volume = {576},
          eid = {A12},
        pages = {A12},
          doi = {10.1051/0004-6361/201425140},
archivePrefix = {arXiv},
       eprint = {1410.2808},
 primaryClass = {astro-ph.CO},
       adsurl = {https://ui.adsabs.harvard.edu/abs/2015A&A...576A..12A}
}

@ARTICLE{2016A&A...586A.122A,
       author = {{Adam}, R. and {Comis}, B. and {Bartalucci}, I. and {Adane}, A. and {Ade}, P. and {Andr{\'e}}, P. and {Arnaud}, M. and {Beelen}, A. and {Belier}, B. and {Beno{\^\i}t}, A. and {Bideaud}, A. and {Billot}, N. and {Bourrion}, O. and {Calvo}, M. and {Catalano}, A. and {Coiffard}, G. and {D'Addabbo}, A. and {D{\'e}sert}, F. -X. and {Doyle}, S. and {Goupy}, J. and {Hasnoun}, B. and {Hermelo}, I. and {Kramer}, C. and {Lagache}, G. and {Leclercq}, S. and {Mac{\'\i}as-P{\'e}rez}, J. -F. and {Martino}, J. and {Mauskopf}, P. and {Mayet}, F. and {Monfardini}, A. and {Pajot}, F. and {Pascale}, E. and {Perotto}, L. and {Pointecouteau}, E. and {Ponthieu}, N. and {Pratt}, G.~W. and {Rev{\'e}ret}, V. and {Ritacco}, A. and {Rodriguez}, L. and {Savini}, G. and {Schuster}, K. and {Sievers}, A. and {Triqueneaux}, S. and {Tucker}, C. and {Zylka}, R.},
        title = "{High angular resolution Sunyaev-Zel'dovich observations of MACS J1423.8+2404 with NIKA: Multiwavelength analysis}",
      journal = {A\& A},
         year = 2016,
        month = feb,
       volume = {586},
          eid = {A122},
        pages = {A122},
          doi = {10.1051/0004-6361/201527616},
archivePrefix = {arXiv},
       eprint = {1510.06674},
 primaryClass = {astro-ph.CO},
       adsurl = {https://ui.adsabs.harvard.edu/abs/2016A&A...586A.122A}
}

@INPROCEEDINGS{2022EPJWC.25700038P,
       author = {{Perotto}, L. and {Adam}, R. and {Ade}, P. and {Ajeddig}, H. and {Andr{\'e}}, P. and {Arnaud}, M. and {Artis}, E. and {Aussel}, H. and {Bartalucci}, I. and {Beelen}, A. and {Beno{\^\i}t}, A. and {Berta}, S. and {Bing}, L. and {Bourrion}, O. and {Calvo}, M. and {Catalano}, A. and {De Petris}, M. and {D{\'e}sert}, F. -X. and {Doyle}, S. and {Driessen}, E.~F.~C. and {Ferragamo}, A. and {Gomez}, A. and {Goupy}, J. and {K{\'e}ruzor{\'e}}, F. and {Kramer}, C. and {Ladjelate}, B. and {Lagache}, G. and {Leclercq}, S. and {Lestrade}, J. -F. and {Mac{\'\i}as-P{\'e}rez}, J. -F. and {Maury}, A. and {Mauskopf}, P. and {Mayet}, F. and {Monfardini}, A. and {Mu{\~n}oz-Echeverr{\'\i}a}, M. and {Paliwal}, A. and {Pisano}, G. and {Pointecouteau}, E. and {Ponthieu}, N. and {Pratt}, G.~W. and {Rev{\'e}ret}, V. and {Rigby}, A.~J. and {Ritacco}, A. and {Romero}, C. and {Roussel}, H. and {Ruppin}, F. and {Schuster}, K. and {Shu}, S. and {Sievers}, A. and {Tucker}, C. and {Yepes}, G.},
        title = "{The NIKA2 Sunyaev-Zeldovich Large Program: Precise galaxy cluster physics for an accurate cluster-based cosmology}",
    booktitle = {mm Universe @ NIKA2 - Observing the mm Universe with the NIKA2 Camera},
         year = 2022,
       series = {European Physical Journal Web of Conferences},
       volume = {257},
        month = jul,
    publisher = {EDP},
          eid = {00038},
        pages = {00038},
          doi = {10.1051/epjconf/202225700038},
archivePrefix = {arXiv},
       eprint = {2111.01729},
 primaryClass = {astro-ph.CO},
       adsurl = {https://ui.adsabs.harvard.edu/abs/2022EPJWC.25700038P}
}

@ARTICLE{Gianfagna2023,
author = {{Gianfagna}, Giulia and {Rasia}, Elena and {Cui}, Weiguang and {De Petris}, Marco and {Yepes}, Gustavo and {Contreras-Santos}, Ana and {Knebe}, Alexander},
title = "{A study of the hydrostatic mass bias dependence and evolution within The Three Hundred clusters}",
journal = {Monthly Notices of the Royal Astronomical Society},
year = 2023,
month = jan,
volume = {518},
number = {3},
pages = {4238-4248},
doi = {10.1093/mnras/stac3364},
archivePrefix = {arXiv},
eprint = {2211.08372},
primaryClass = {astro-ph.CO},
adsurl = {https://ui.adsabs.harvard.edu/abs/2023MNRAS.518.4238G}
}

@ARTICLE{Mostoghiu2019,
author = {{Mostoghiu}, Robert and {Knebe}, Alexander and {Cui}, Weiguang and {Pearce}, Frazer R. and {Yepes}, Gustavo and {Power}, Chris and {Dave}, Romeel and {Arth}, Alexander},
title = "{The Three Hundred Project: The evolution of galaxy cluster density profiles}",
journal = {Monthly Notices of the Royal Astronomical Society},
year = 2019,
month = mar,
volume = {483},
number = {3},
pages = {3390-3403},
doi = {10.1093/mnras/sty3306},
archivePrefix = {arXiv},
eprint = {1812.04009},
primaryClass = {astro-ph.GA},
adsurl = {https://ui.adsabs.harvard.edu/abs/2019MNRAS.483.3390M}
}

@ARTICLE{2013ApJ...778...52S,
       author = {{Sayers}, J. and {Mroczkowski}, T. and {Zemcov}, M. and {Korngut}, P.~M. and {Bock}, J. and {Bulbul}, E. and {Czakon}, N.~G. and {Egami}, E. and {Golwala}, S.~R. and {Koch}, P.~M. and {Lin}, K. -Y. and {Mantz}, A. and {Molnar}, S.~M. and {Moustakas}, L. and {Pierpaoli}, E. and {Rawle}, T.~D. and {Reese}, E.~D. and {Rex}, M. and {Shitanishi}, J.~A. and {Siegel}, S. and {Umetsu}, K.},
        title = "{A Measurement of the Kinetic Sunyaev-Zel'dovich Signal Toward MACS J0717.5+3745}",
      journal = {ApJ},
         year = 2013,
        month = nov,
       volume = {778},
       number = {1},
          eid = {52},
        pages = {52},
          doi = {10.1088/0004-637X/778/1/52},
archivePrefix = {arXiv},
       eprint = {1312.3680},
 primaryClass = {astro-ph.CO},
       adsurl = {https://ui.adsabs.harvard.edu/abs/2013ApJ...778...52S}
}

@ARTICLE{2012ApJ...761...47M,
       author = {{Mroczkowski}, Tony and {Dicker}, Simon and {Sayers}, Jack and {Reese}, Erik D. and {Mason}, Brian and {Czakon}, Nicole and {Romero}, Charles and {Young}, Alexander and {Devlin}, Mark and {Golwala}, Sunil and {Korngut}, Phillip and {Sarazin}, Craig and {Bock}, James and {Koch}, Patrick M. and {Lin}, Kai-Yang and {Molnar}, Sandor M. and {Pierpaoli}, Elena and {Umetsu}, Keiichi and {Zemcov}, Michael},
        title = "{A Multi-wavelength Study of the Sunyaev-Zel'dovich Effect in the Triple-merger Cluster MACS J0717.5+3745 with MUSTANG and Bolocam}",
      journal = {ApJ},
         year = 2012,
        month = dec,
       volume = {761},
       number = {1},
          eid = {47},
        pages = {47},
          doi = {10.1088/0004-637X/761/1/47},
archivePrefix = {arXiv},
       eprint = {1205.0052},
 primaryClass = {astro-ph.CO},
       adsurl = {https://ui.adsabs.harvard.edu/abs/2012ApJ...761...47M}
}

@ARTICLE{2005RvMP...77..207V,
       author = {{Voit}, G. Mark},
        title = "{Tracing cosmic evolution with clusters of galaxies}",
      journal = {Reviews of Modern Physics},
         year = 2005,
        month = apr,
       volume = {77},
       number = {1},
        pages = {207-258},
          doi = {10.1103/RevModPhys.77.207},
archivePrefix = {arXiv},
       eprint = {astro-ph/0410173},
 primaryClass = {astro-ph},
       adsurl = {https://ui.adsabs.harvard.edu/abs/2005RvMP...77..207V}
}

@ARTICLE{2009MNRAS.394..479A,
       author = {{Ameglio}, S. and {Borgani}, S. and {Pierpaoli}, E. and {Dolag}, K. and {Ettori}, S. and {Morandi}, A.},
        title = "{Reconstructing mass profiles of simulated galaxy clusters by combining Sunyaev-Zeldovich and X-ray images}",
      journal = {MNRAS},
         year = 2009,
        month = mar,
       volume = {394},
       number = {1},
        pages = {479-490},
          doi = {10.1111/j.1365-2966.2008.14324.x},
       adsurl = {https://ui.adsabs.harvard.edu/abs/2009MNRAS.394..479A}
}

@ARTICLE{2007MNRAS.382..397A,
       author = {{Ameglio}, S. and {Borgani}, S. and {Pierpaoli}, E. and {Dolag}, K.},
        title = "{Joint deprojection of Sunyaev-Zeldovich and X-ray images of galaxy clusters}",
      journal = {MNRAS},
         year = 2007,
        month = nov,
       volume = {382},
       number = {1},
        pages = {397-411},
          doi = {10.1111/j.1365-2966.2007.12384.x},
archivePrefix = {arXiv},
       eprint = {0708.3737},
 primaryClass = {astro-ph},
       adsurl = {https://ui.adsabs.harvard.edu/abs/2007MNRAS.382..397A}
}

@ARTICLE{2013JCAP...07..008H,
       author = {{Hasselfield}, Matthew and {Hilton}, Matt and {Marriage}, Tobias A. and {Addison}, Graeme E. and {Barrientos}, L. Felipe and {Battaglia}, Nicholas and {Battistelli}, Elia S. and {Bond}, J. Richard and {Crichton}, Devin and {Das}, Sudeep and {Devlin}, Mark J. and {Dicker}, Simon R. and {Dunkley}, Joanna and {D{\"u}nner}, Rolando and {Fowler}, Joseph W. and {Gralla}, Megan B. and {Hajian}, Amir and {Halpern}, Mark and {Hincks}, Adam D. and {Hlozek}, Ren{\'e}e and {Hughes}, John P. and {Infante}, Leopoldo and {Irwin}, Kent D. and {Kosowsky}, Arthur and {Marsden}, Danica and {Menanteau}, Felipe and {Moodley}, Kavilan and {Niemack}, Michael D. and {Nolta}, Michael R. and {Page}, Lyman A. and {Partridge}, Bruce and {Reese}, Erik D. and {Schmitt}, Benjamin L. and {Sehgal}, Neelima and {Sherwin}, Blake D. and {Sievers}, Jon and {Sif{\'o}n}, Crist{\'o}bal and {Spergel}, David N. and {Staggs}, Suzanne T. and {Swetz}, Daniel S. and {Switzer}, Eric R. and {Thornton}, Robert and {Trac}, Hy and {Wollack}, Edward J.},
        title = "{The Atacama Cosmology Telescope: Sunyaev-Zel'dovich selected galaxy clusters at 148 GHz from three seasons of data}",
      journal = {JCAP},
         year = 2013,
        month = jul,
       volume = {2013},
       number = {7},
          eid = {008},
        pages = {008},
          doi = {10.1088/1475-7516/2013/07/008},
archivePrefix = {arXiv},
       eprint = {1301.0816},
 primaryClass = {astro-ph.CO},
       adsurl = {https://ui.adsabs.harvard.edu/abs/2013JCAP...07..008H}
}

@ARTICLE{2024A&A...682A.147M,
       author = {{Mu{\~n}oz-Echeverr{\'\i}a}, M. and {Mac{\'\i}as-P{\'e}rez}, J.~F. and {Pratt}, G.~W. and {Pointecouteau}, E. and {Bartalucci}, I. and {De Petris}, M. and {Ferragamo}, A. and {Hanser}, C. and {K{\'e}ruzor{\'e}}, F. and {Mayet}, F. and {Moyer-Anin}, A. and {Paliwal}, A. and {Perotto}, L. and {Yepes}, G.},
        title = "{The hydrostatic-to-lensing mass bias from resolved X-ray and optical-IR data}",
      journal = {\aap},
         year = 2024,
        month = feb,
       volume = {682},
          eid = {A147},
        pages = {A147},
          doi = {10.1051/0004-6361/202347584},
archivePrefix = {arXiv},
       eprint = {2312.01154},
 primaryClass = {astro-ph.CO},
       adsurl = {https://ui.adsabs.harvard.edu/abs/2024A&A...682A.147M}
}

\appendix
\section{GHP effective length maps compared to $n_e$-inferred ones}
\label{appendix:A}

\begin{figure}
    \centering
    \includegraphics[width=\linewidth, clip, trim={0 0 0 0}]{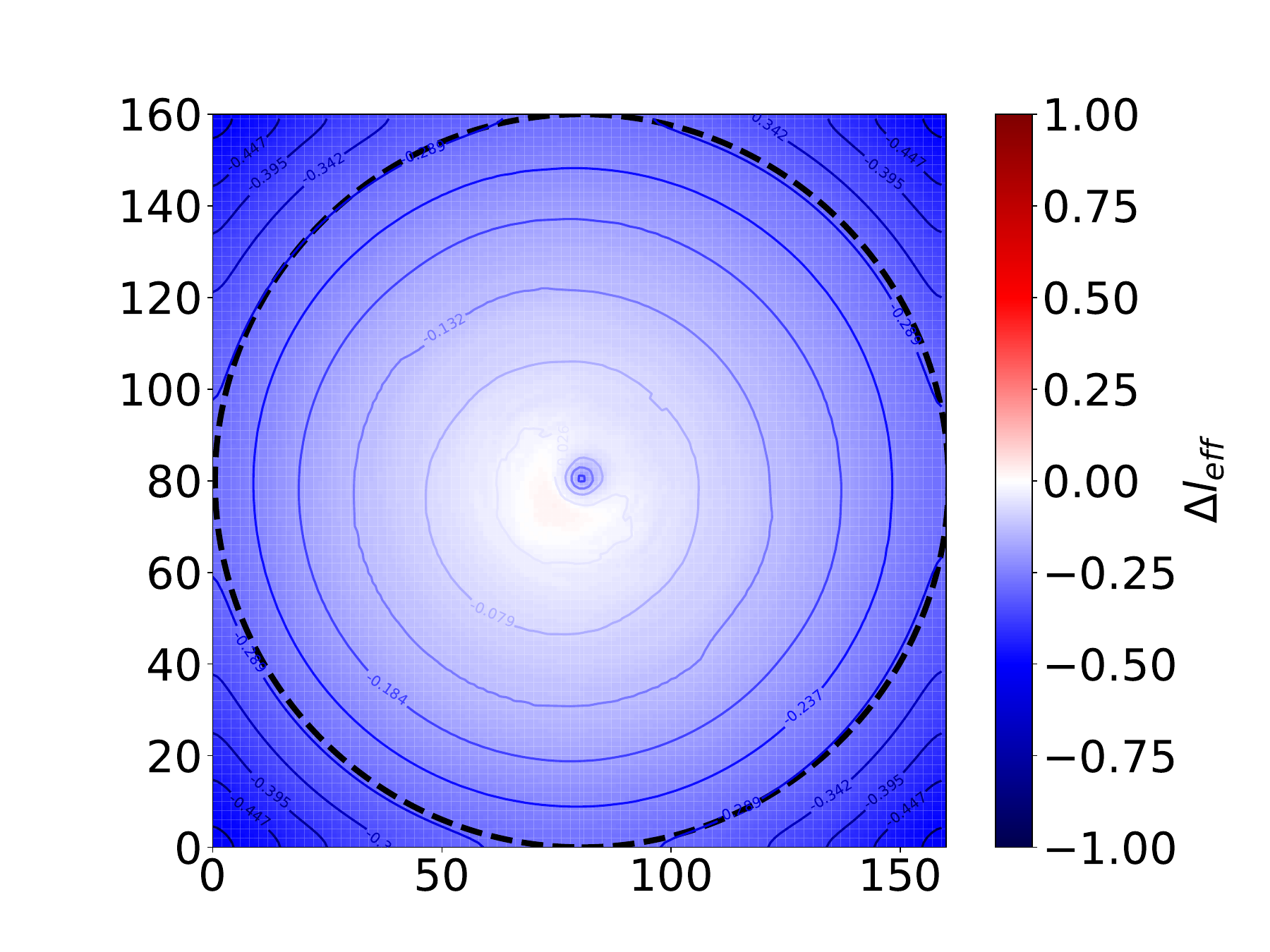}
    \caption{Stacked comparative map of the GHP-inferred and Vikhlinin-inferred effective length maps. The dashed line represents the spatial extension of the map, covering $\rth$ for each cluster. The contours span the range $\pm 50\%$ while the colorbar spans the range $\pm 100\%$.}
    \label{fig:GHP_v_Vik_leffmaps}
\end{figure}

We show here the shape comparison between different effective length templates, obtained using the GHP model (defined in Eq. \ref{eq:GHP_def}) and using the Vikhlinin model (defined in Eq. \ref{eq:vikhlinin_model}). 
As the Vikhlinin model gives the most precise estimation of $\leff$ out of the two models based on the fitting of the density profile (Sect. \ref{subsec:GHP_v_neinferred}), we do not consider the $\beta$ model here.

For each cluster the GHP effective length map is obtained following one of the three calibrations of the GHP model (Relaxed, Disturbed or Median), while the Vikhlinin effective length is obtained by directly fitting the density profile for each cluster.
We show below the stacking S of all the bias maps between the GHP and the Vikhlinin-inferred effective length maps, i.e.
\begin{equation}
    S = \mathrm{med}\left(\frac{l_{eff, GHP} - l_{eff, Vik}}{l_{eff, GHP}}\right).
\end{equation}
We show that on average, the effective lengths recovered using the Vikhlinin model are higher than those obtained using the GHP model. 
This effect is particularly strong in the outskirts of the clusters at radii $\gtrsim \rth$, as the Vikhlinin model cannot replicate the flattening of $\leff$, and in the innermost regions, with a strong peak towards low $\leff$ values in the GHP model that cannot be reproduced with the Vikhlinin model.
The Vikhlinin and GHP effective lengths show agreement within $10\%$ only in the inner regions of the cluster below $\sim 0.5 \rth$, excluding the central peak.
As discussed in the bulk of the paper, this effect is a signature of the gas clumping, poorly recovered when fitting 1D density profiles.
This also explains the underestimated $\tszx$ values that we obtained when using $\leff$ maps derived from density profiles (see the discussion in Sect. \ref{subsec:GHP_v_neinferred}).
\end{document}